\documentclass[11pt]{article}
\usepackage{amssymb,amsmath,amsfonts}
\usepackage{graphicx}
\usepackage{graphics}
\usepackage{eepic,epsfig}

1.60cm \makeatletter \@addtoreset{equation}{section}

\makeatother
\begin{document}

\title{Fermionic vacuum polarization by an Abelian magnetic tube in the cosmic string spacetime}
\author{M. S. Maior de Sousa \thanks{E-mail: kaelsousa@gmail.com}\ ,  R. Freire Ribeiro \thanks{E-mail: 
rfreire@fisica.ufpb.br} and  E. R. Bezerra de Mello \thanks{E-mail: emello@fisica.ufpb.br} \
\\
Departamento de F\'{\i}sica-CCEN\\
Universidade Federal da Para\'{\i}ba\\
58.059-970, J. Pessoa, PB\\
C. Postal 5.008\\
Brazil}

\maketitle

\begin{abstract}
In this paper, we consider a charged massive fermionic quantum field  in the idealized cosmic string spacetime and in the presence of a magnetic field confined in a cylindrical tube of finite radius. Three distinct configurations
for the magnetic fields are taken into account: (i) a cylindrical shell of radius $a$, (ii) a magnetic field proportional to $1/r$ and (iii) a constant magnetic field. In these three cases, the axis of the infinitely long
tube of radius $a$ coincides with the cosmic string. Our main objectives in this paper are to analyze the fermionic condensat (FC) e and the vacuum expectation value (VEV) of the fermionic energy-momentum tensor. In order to do that, we explicitly construct the complete set of normalized wave-functions for each configuration of magnetic field. We show that in the region outside the tube, the FC and the VEV of the energy-momentum tensor are decomposed into two parts: the first ones correspond to the zero-thickness magnetic flux contributions, and the seconds are induced by the non-trivial structure of the magnetic field, named core-induced contributions. The latter present specific forms depending on the magnetic field configuration considered. We also show that the VEV of the energy-momentum tensor is diagonal, obeys the conservation condition  and its trace is expressed in terms of the fermionic condensate. The zero-thickness contributions to the FC and VEV of the energy-momentum tensor, depend only on the fractional part of the ration of the magnetic  flux inside the tube by the quantum one. As to the core-induced contributions they depend on the  total magnetic flux inside the tube, and consequently, in general, are not a periodic function of the magnetic flux. 
\end{abstract}

\bigskip


\bigskip

\section{Introduction}
\label{Int}

According to the Big Bang theory, the Universe has been expanding and cooling. During its expansion, the spontaneous breaking of fundamental symmetries leads the universe to a series of phase transitions. In most interesting model of high-energy physics, the formation of topological defects such as domain  walls, monopoles, cosmic string, among others are predicted  to occur \cite{vilenkin}. These topologically stable structures have a number of interesting observable consequences, the detection of which would provide an important link between cosmology and particle physics. 

The cosmic strings are among the various type of topological defects, the most studied. Though the recent observations of the cosmic microwave background radiation have ruled out them as the primary source for primordial density perturbations, cosmic strings give rise to a number of interesting physical effects such as gamma rays bursts \cite{Bere}, the emission of gravitational waves \cite{Damour} and the generation of high-energy cosmic rays \cite{sigl}. String-like defects also appear in a number of condensed matter systems, including liquid crystals and graphene made structures.

At large distances, the spacetime geometry for an ideal infinitely straight cosmic string, has a conical topology with a planar angle deficit proportional to its linear mass density. In quantum field theory the corresponding non-trivial topology induces nonzero vacuum expectation values (VEVs) for physical observables. These vacuum polarization effects in quantum field theory induced by this conical structure have been considered in a large number of papers. In the specific analysis for the VEV of the energy-momentum tensor, explicit calculations have been developed for scalar, fermionic, and electromagnetic fields \cite{TMDA}-\cite{Bezerra}. The Casimir-Polder forces acting on a polarizable microparticle in the geometry of a cosmic string have been investigated in \cite{saharian1}-\cite{mello3}. For charged fields, considering the presence of a magnetic flux running along the cosmic strings, there appear additional contributions to the corresponding vacuum polarization effects \cite{dowker2}, \cite{guima94}-\cite{jean08}. The magnetic flux along the cosmic string also induces vacuum current densities. This phenomenon has been investigated for scalar fields in \cite{sriram}-\cite{yua}. The analysis of the induced fermionic currents in higher dimensional cosmic string spacetime in the presence of a magnetic flux has been developed in \cite{mello2010}. In these analysis the authors have shown that induced vacuum current densities along the azimuthal direction appear if the ratio of the magnetic flux by the quantum one has a nonzero fractional part. In \cite{Mello13} and \cite{saharian14} it was investigated the induced current and the vacuum polarization effects, respectively, associated with quantum charged massive fermionic field in the geometry of a compactified cosmic string spacetime,  considering  the presence of an infinitely thin magnetic field running along the string's axis. The induced scalar current in the compactified cosmic string spacetime, has been developed in \cite{Braganca}. Moreover, the fermionic current induced by a magnetic flux in a (2 + 1)-dimensional conical spacetime and in the presence of a circular boundary has also been analyzed in \cite{mello10.2}. In \cite{mikael} the authors have analyzed the VEV of the azimuthal fermionic current, and showed that this current can be expressed as a sum  of two distinct contributions: the first one is due to the presence of an idealized cosmic string carrying a thin magnetic field, that depends only on the fractional part of the ratio of the total magnetic flux by the quantum one, and the second contribution is induced by the non-vanishing core of the magnetic field. The latter, in general, is not a periodic function function of the total  magnetic flux.

 Many of the publications involving the analysis of the VEV of the energy-momentum tensor around a cosmic string deal mainly with the case of the idealized geometry where the string is assumed to have zero thickness. However, realistic cosmic strings have inner structure, characterized by a finite width core determined by the symmetry breaking scale at which it is formed. So charged fields interact not only with the magnetic field, but also with the core. Specifically for fermionic fields this can lead to interesting physics like as fermionic zero modes. 
However, as it was pointed out in \cite{mikael}, there are no analytical solutions for the system corresponding to the Abelian Nielsen and Olesen vortex carrying a magnetic flux, which is the system that we want to investigate. Only numerical solutions can be obtained (See references there). So, in order to provide some improvement in the calculations of the fermionic current, in \cite{mikael} we have adopted a simplified model for the cosmic string discarding its inner structure and considering a non-vanishing extension to the magnetic field. 
In general, the radii of the string core and of the gauge field flux can be different. For the Abelian Higgs model, these radii are determined by the inverse of 
Higgs mass, $1/m_H$, and the inverse of the vector mass, $1/m_v$, respectively. 
Assuming that $m_H>m_v$, and that the energies associated with the fermions are smaller than $m_H$, the inner structure of the string can be ignored in a first approximation. This is the main essence of our work. This fact can be implemented by admitting that the fermionic wave function obeys the Dirichlet boundary condition at the top of the string, i.e.,  $\psi=0$ at $r=0$. This same boundary condition will be adopted in this present paper. \footnote{The analysis of VEV of scalar azimuthal current around a cosmic string considering a general cylindrically symmetric inner structure to it has been developed in Ref. \cite{Aram15}; however explicit expressions were only obtained by assuming specific models for the string's core.}

Here, in this paper we decided to continue in a similar line of previous investigation of Ref. \cite{mikael},  considering the fermionic condensate (FC) and the VEV of the energy-momentum tensor associated with a massive charged fermionic field in the presence of an ideal cosmic string surrounded by a cylindrical magnetic tube that can be taken in three different configurations: $(i)$ a cylindrical shell of magnetic field, $(ii)$ a magnetic field decaying with $1/r$ and $(iii)$ a cylindrical homogeneous magnetic field with radius $a$. For all these cases the axis of the magnetic tube coincides with the cosmic string.  We want to study how the topological effect and the presence of the magnetic tube modify the above mentioned quantities. These VEVs carry important informations about the global properties of the background spacetime and also about the configurations of the magnetic fields. Specifically the FC plays an important role in models of dynamical breaking of the chiral symmetry. As to the VEV of the energy-momentum tensor, in addition to describe the physical structure of the quantum field at a given point, it acts as a source of gravity in the semiclassical Einstein equations and is of relevant importance in modeling self-consistent dynamics involving fermionic fields.

This paper is organized as follow. In the next section we present the geometry background associated with the spacetime under consideration and provide the complete set of normalized fermionic positive- and negative-energy wave function. In section \ref{fermicond}, by using the mode summation formula, we evaluate the fermionic condensate (FC) in a compact form. We show that FC can be written as the sum of a contribution due to a infinitely thin magnetic field plus a core-induced contribution. Specifically we analyze the core-induced part. We present its  behavior in some specific regions, $r\gtrsim a$ and $r>>a$, and  provide some plots that exhibit its explicit dependences with some physical relevant parameters. 
 In section \ref{EMT}, also by the using the mode sum method, we explicitly evaluate all components of the energy-momentum tensor (EMT), and show that they can be written by the sum of two contributions as it was shown in previous analysis. Moreover, we show that the EMT is written in a diagonal form. The main objective of this section is devoted to analyze the core-induced contribution. We prove that, in similar way as the contribution due to the infinitely thin magnetic field, the core-induced contribution obeys the conservation condition and the trace identity. Also, in section \ref{EMT}, we present the behavior of the EMT in some specific regions of the space and provide some plots that exhibit their behavior as function of the total magnetic field and the parameter that codify the conical structure of the spacetime. Finally in section \ref{Concl} we present our most relevant conclusions about our investigations. Throughout the paper we use the units with $G=\hbar=c=1$.

\section{The geometry background and the Dirac wave functions}
\label{Geometry}

In the presence of an external electromagnetic field, $A_\mu$, the quantum dynamic of a massive charged fermionic field in curved spacetime is described by the Dirac equation,
\begin{equation}
\label{2.2}
i\gamma^{\mu} (\nabla_\mu + ieA_\mu)\psi-m\psi=0, \ \ \nabla_\mu=\partial_\mu + \Gamma_\mu \ ,
\end{equation}
where $\gamma^\mu$ represents the Dirac matrices in curved spacetime and $\Gamma_\mu$ the spin connection. Both matrices are given in terms of the flat spacetime Dirac matrices, $\gamma^{(a)}$, by the relations
\begin{equation}
\label{2.3}
\Gamma_\mu=-\frac{1}{4}\gamma^{(a)}\gamma^{(b)}e^\nu_{(a)}e_{(b)\nu;\mu}, \ \ \gamma^\mu = {e^\mu}_{(a)}\gamma^{(a)} \ ,
\end{equation}
being ${e^\mu}_{(a)}$ the basis tetrad that satisfies the relation ${e^\mu}_{(a)}{e^\nu}_{(b)}\eta^{ab}=g^{\mu\nu}$, with $\eta^{ab}$ the Minkowski spacetime metric tensor.

The geometry associated with an idealized cosmic string along the $z$-axis, can be expressed in cylindrical coordinates, by the line element below:
\begin{equation}
\label{2.1}
ds^2=dt^2 - dr^2 - r^2d\phi^2-dz^2 \ .
\end{equation}
In the above expression the coordinate system take values in range $r\geq 0$, $0\leq\phi\leq\phi_0=2\pi/q$ and $-\infty\leq (t, \ z)\leq+\infty$. The parameter $q$ is related with the planar angle deficit that can be given in terms of the mass per unit
length of the string, $\mu_0$,  by the relation $q^{-1}=1-4\mu_0$.

The system that we want to analyze consists of a charged fermionic quantum field in the cosmic string background, taking into account the presence of three different configurations of magnetic fields. They are: $(i)$ A magnetic field concentrated on a cylindrical shell of radius $a$, $(ii)$ a magnetic field proportional to $1/r$, and finally $(iii)$ a homogeneous magnetic field inside the tube. In these three cases, the axis of the infinitely long tube of radius $a$ coincides with the cosmic string. In what follows we will represent these configurations of magnetic fields by a four-vector potential, $A_\mu$, in a compact form below:
\begin{equation}
\label{2.01}
A_{\mu}=(0,0,A_\phi(r),0) \ ,
\end{equation}
with
\begin{equation}
\label{2.012}
A_\phi(r)=-\frac{q\Phi}{2\pi}a(r) \ .
\end{equation}
For the first model,
\begin{equation}
\label{2.013}
a(r)=\Theta(a-r) \ .
\end{equation}
For the second and third models, we can represent the radial function $a(r)$ by:
\begin{equation}
\label{2.014}
a(r)=f(r)\Theta (a-r)+\Theta (r-a) \ ,
\end{equation}
with
\begin{eqnarray}
\label{2.015}
f(r)=\left\{\begin{array}{cc}
r/a,&\mbox{for the second model} \  ,  \\
r^2/a^2,&\mbox{for the third model} \  .
\end{array}
\right.
\end{eqnarray}
In the previous expressions, $\Theta(z)$ represents the Heaviside function, and $\Phi$ the total magnetic flux.

In the geometry described by \eqref{2.1} the Dirac gamma matrices can be expressed as \cite{mikael},
\begin{equation}
\label{2.6}
\gamma^0=\gamma^{(0)}=\left( \begin{array}{cc}
1 & 0 \\
0 & -1 \end{array} \right), \ \ \gamma^{l}=\left( \begin{array}{cc}
0 & \sigma^l \\
-\sigma^l & 0 \end{array} \right),
\end{equation}
where we have introduced the $2 \times 2$ matrices for $l=(r, \ \phi, \ z)$:
\begin{equation}
\label{2.7}
\sigma^{r}=\left( \begin{array}{cc}
0 & e^{-iq\phi} \\
e^{iq\phi} & 0 \end{array} \right), \ \sigma^{\phi}=-\frac{i}{r}\left( \begin{array}{cc}
0 & e^{-iq\phi} \\
-e^{iq\phi} & 0 \end{array} \right), \ \sigma^{z}=\left( \begin{array}{cc}
1 & 0 \\
0 & -1 \end{array} \right).
\end{equation}
For this case the spin connection and combination appearing in the Dirac equation we have,
\begin{equation}
\label{2.8}
\Gamma_\mu=\frac{1-q}{2}\gamma^{(1)}\gamma^{(2)}\delta^\phi_\mu, \ \ \ \gamma^\mu \Gamma_\mu=\frac{1-q}{2r}\gamma^r.
\end{equation}
So, the Dirac equation take the form
\begin{equation}
\label{2.9}
\left(\gamma^\mu(\partial_\mu + ieA_\mu)+\frac{1-q}{2r}\gamma^r + im\right)\psi=0 \  .
\end{equation}
	
In the present paper we are interested to investigate the effects associated with the non-vanishing structure of the magnetic tube of radius $a$, on the fermionc condensate and on the vacuum expectation value of the fermionic energy-momentum tensor. These calculations will be developed by using the summation method over the normalized fermionic modes. In recent publucation \cite{mikael} the authors have presented the complete procedure to construct the positive- and negative-energy fermionic wavefunctions in the region outside the magnetic tube, $r\geq a$, considering three different configurations of magnetic fields. The positive- and negative-energy wavefunctions are specified by the complete set of quantum numbers $\sigma=(\lambda, \ k, \ j, \ s)$. These functions can be written as show below:	
\begin{equation}
\label{psi-out}
\psi^{(\pm)}_{\sigma(out)}(x)=C^{(\pm)}_{(out)}e^{\mp i(Et-kz)}e^{iq(j-1/2) \phi}\left( \begin{array}{c}
g_{\beta_j}(\lambda a, \lambda r) \ ,  \\
\pm i\epsilon_j \rho_s b^{(\pm)}_s g_{\beta_j+\epsilon_j}(\lambda a, \lambda r) e^{iq\phi} \\
\rho_s g_{\beta_j}(\lambda a, \lambda r) \\
\mp i\epsilon_j b^{(\pm)}_s g_{\beta_j+\epsilon_j}(\lambda a, \lambda r) e^{iq\phi} \end{array} \right),
\end{equation}
where we have introduced the notations,
\begin{equation}
\label{g-beta}
g_{\beta_j}(\lambda a, \lambda r)=\frac{\tilde{Y}_{\beta_j}(\lambda a)J_{\beta_j}(\lambda r)
-\tilde{J}_{\beta_j}(\lambda a)Y_{\beta_j}(\lambda r)}{\sqrt{(\tilde{Y}_{\beta_j}(\lambda a))^2
+(\tilde{J}_{\beta_j}(\lambda a))^2}}  \  ,
\end{equation}
\begin{equation}
\label{g-beta1}
g_{\beta_j+\epsilon_j}(\lambda a, \lambda r)=\frac{\tilde{Y}_{\beta_j}(\lambda a)
J_{\beta_j+\epsilon_j}(\lambda r)-\tilde{J}_{\beta_j}(\lambda a)Y_{\beta_j+\epsilon_j}
(\lambda r)}{\sqrt{(\tilde{Y}_{\beta_j}(\lambda a))^2+(\tilde{J}_{\beta_j}(\lambda a))^2}}  \   ,
\end{equation}
\begin{eqnarray}
\label{rho}
\rho_s=\frac{E+s\sqrt{\lambda^2 + m^2}}{k} \ , \  s= \pm 1  
\end{eqnarray}
and	
\begin{equation}
\label{2.26}
b^{(\pm)}_s=\frac{\pm m+s\sqrt{\lambda^2+m^2}}{\lambda} \  .
\end{equation}

 We can see, by the general structure of the wavefunctions above, that they are eigenfunctions of the total angular moment along the $z-$axis, as show below:
\begin{eqnarray}
{\hat{J}}\psi^{(\pm)}=\left(-i\partial_\phi+i\frac q2\gamma^{(1)}\gamma^{(2)}\right)\psi^{(\pm)}=qj \psi^{(\pm)} \  ,
\end{eqnarray}
with eigenvalues, 
\begin{eqnarray}
j=n+\frac12 \ , \ n=0, \ \pm 1, \ \pm2, \  ... 
\end{eqnarray}
Moreover, the expressions \eqref{g-beta} and \eqref{g-beta1} are defined according to, 
\begin{eqnarray}
\label{Z.Bessel}
{\tilde{Z}}_{\beta_j}(z)=\epsilon_jZ_{\beta_j+\epsilon_j}(z)-{\cal{V}}_j^{(i)}(\lambda, a)
Z_{\beta_j} (z)  \ , \  {\rm with} \
{\cal{V}}^{(i)}_j(\lambda, a)=\frac{R_2^{(i)}(\lambda, a)}{R^{(i)}_1(\lambda, a)} \ ,
\end{eqnarray}
where $Z_\nu$ represents the Bessel, $J_\mu$, or Newmann, $Y_\mu$, functions, and $R^{(i)}_1(\lambda, r)$ and $R_2^{(i)}(\lambda, r)$ the components of the two independent radial functions in the region inside the tube according to the general expression below:
\begin{equation}
\label{in}
\psi^{(\pm)}_{i(in)}(x)=C^{(\pm)}_{(in)} e^{\mp i(Et-kz)}e^{iq(j-1/2)\phi}\left( \begin{array}{c}
R^{(i)}_1(\lambda, r)\\
\pm i\rho_s b^{(\pm)}_{s} R^{(i)}_2(\lambda, r) e^{iq\phi} \\
\rho_s R^{(i)}_1(\lambda, r) \\
\mp i b^{(\pm)}_{s}R^{(i)}_2(\lambda, r) e^{iq\phi} \end{array} \right) \   .
\end{equation}
In the wave-functions we consider $s=\pm 1$, $\lambda \geq 0$, $k\in(-\infty, \infty)$. The orders of the Bessel and Newmann functions are expressed as $\beta_j$ and $\beta_j+\epsilon_j$, defined below:
\begin{eqnarray}
\beta_j=q|j+\alpha|-\frac{\epsilon_j}{2} \ , \ {\rm with} \ \alpha=-\frac{q\Phi}{2\pi} \  , 
\end{eqnarray}
being $\epsilon_j=1$ for $j+\alpha>0$ and $\epsilon_j=-1$ for $j+\alpha<0$. Finally the energy is given by,
\begin{equation}
E=\sqrt{\lambda^2+k^2+m^2} \  .
\end{equation}

The normalization constants for the positive- and negative-energy wavefunction,  \eqref{psi-out}, can be obtained form the normalization condition,
\begin{equation}
\label{Ren}
\int{d^3x\sqrt{-g^{(3)}}\left(\psi^{(\pm)}_\sigma\right)^\dagger\psi^{(\pm)}_{\sigma'}}=
\delta_{\sigma,\sigma'} \ ,
\end{equation}
where delta symbol on the right-hand side is understood as the Dirac delta function for continuous quantum numbers $\lambda \ \mbox{and} \ k$, and as the Kronecker delta for discrete ones $n, \ j\ \mbox{and} \ s$, and $g^{(3)}$ is the determinant of the spatial metric tensor. The integral over the radial coordinate should be done in the interval $[0, \ \infty)$. After some intermediate steps we found: 
\begin{equation}
\label{Ren1}
C^{(\pm)}_{(out)}=\frac{1}{2\pi}\left[\frac{q\lambda}{(1+\rho^2_s)\left(1+(b^{(\pm)}_s)^2\right)}\right]^{1/2}  \  ,
\end{equation}

Having found the complete set of outside normalized wavefunctions we are able to calculate the FC and the VEV of the energy-momentum tensor. In fact our main objective in this paper is to evaluate the contributions on these observables induced by the non-vanishing core of the magnetic tube. The fermionic condensate is defined
as the VEV of the scalar density ${\bar{\Psi}\Psi}$, $\langle 0| {\bar{\Psi}\Psi}|0 \rangle\equiv\langle {\bar{\Psi}\Psi} \rangle$, where $|0\rangle$ corresponds to the fermionic vacuum and ${\bar{\Psi}}=\Psi^\dagger\gamma^{(0)}$ is the Dirac adjoint. As to the VEV of the energy-momentum tensor, $\langle 0| T_{\mu\nu}|0\rangle\equiv \langle T_{\mu\nu}\rangle$. Expanding the field operator in terms of the complete set $\{\psi_{\sigma(out)}^{(+)}, \ \psi_{\sigma(out)}^{(-)}\}$. These analysis will be developed in the next two sections.

\section{Fermionic condensate}
\label{fermicond}

Here we shall evaluate the fermionic condensate by using the mode sum formula,
\begin{equation}
\label{condesate}
\left\langle\bar{\Psi}\Psi\right\rangle=\sum_{\sigma}{\bar{\psi}^{(-)}_\sigma (x){\psi}^{(-)}_\sigma (x)}  \    ,
\end{equation}
where the summation over the set $\sigma$ corresponds to
\begin{equation}
\label{sum}
\sum_{\sigma}=\int^{\infty}_{0} d\lambda \int^{+\infty}_{-\infty} dk \sum_{j=\pm 1/2, ...} \sum_{s=\pm 1}  \  .
\end{equation}

Using the explicit expression for the outside negative-energy wave-function defined in (\ref{psi-out}) and \eqref{Ren1},  we can show that the contributions for $s=1$ and $s=-1$ give the same result, consequently the fermionic condensate has the following form:
\begin{equation}
\label{condesate1}
\left\langle\bar{\Psi}\Psi\right\rangle = -\frac{q m}{(2\pi)^2}\sum_{j}{\int_{-\infty}^{\infty}{dk\int_{0}^{\infty}{d\lambda\frac{\lambda}{E}\left(g_{\tilde{\beta}_j}^{2}(\lambda a, \lambda r)+g_{\beta_j}^{2}(\lambda a, \lambda r)\right)}}} \  , 
\end{equation}
where ${\tilde{\beta}}_j={\beta}_j+\epsilon_j$.

Taking into account \eqref{Z.Bessel} and using the relations between the Bessel and Hankel functions \cite{Abramo}, we find that,
\begin{equation}
\label{condesate2}
\left\langle\bar{\Psi}\Psi\right\rangle = \left\langle\bar{\Psi}\Psi\right\rangle_{s} + \left\langle\bar{\Psi}\Psi\right\rangle_{c} .
\end{equation}
Where we have shown that the fermionic condensate can be expressed as the sum of two different contributions: the first one, $\left\langle\bar{\Psi} \Psi\right\rangle_{s}$, corresponds to the contribution due to line of magnetic flux running along the string's axis, and the second one, $\left\langle\bar{\Psi} \Psi\right\rangle_{c}$, takes into account the non-vanishing structure of the magnetic tube. The latter presents different expressions for different configurations of the magnetic field. Their explicit forms are presented bellow
\begin{equation}
\label{condesate3}
\left\langle\bar{\Psi}\Psi\right\rangle_{s} = -\frac{q m}{(2\pi)^2}\sum_{j}{\int_{-\infty}^{\infty}{dk\int_{0}^{\infty}{d\lambda\frac{\lambda}{E}\left(J_{\tilde{\beta}_j}^{2}(\lambda a, \lambda r)+J_{\beta_j}^{2}(\lambda a, \lambda r)\right)}}}
\end{equation}
and
\begin{equation}
\label{condesate4}
\left\langle\bar{\Psi}\Psi\right\rangle_{c} = \frac{q m}{2(2\pi)^2}\int^{\infty}_{-\infty}{dk}\sum_{j}\int^{\infty}_{0}
{d\lambda \frac{\lambda}{E} \tilde{J}_{\beta_j}(\lambda a) \sum^{2}_{l=1}{\frac{(H^{(l)}_{\beta_j}
(\lambda r))^2+(H^{(l)}_{\tilde{\beta}_j}(\lambda r))^2}{\tilde{H}^{(l)}_{\beta_j}(\lambda a)}}}  \ ,  
\end{equation}
where the notation $\tilde{\beta_j}=\beta_j + \epsilon_j$ was introduced and the $H^{(l)}_{\nu}(x)$ with $l=1, \ 2$ represents the Hankel function.

At this point we would like to analyze separately both contributions.

\subsection{Fermionic condensate induced by the zero-thickness magnetic flux}
\label{FC_0} 
The FC induced by the zero-thickness magnetic flux in an idealized cosmic string spacetime, has been  given by \eqref{condesate3}. This explicit calculation was developed in \cite{saharian14}. Here we briefly review its more relevant results. 
Substituting the identity 
\begin{equation}
\frac{1}{\sqrt{m^{2}+k^{2}+\lambda ^{2}}}=\frac{2}{\sqrt{\pi }}%
\int_{0}^{\infty }ds\ e^{-(m^{2}+k^{2}+\lambda ^{2})s^{2}} \  \label{ident1}
\end{equation}
into \eqref{condesate3}, we can integrate over the variable $k$. As to
the integral over $\lambda$ we use \cite{Grad} and we get,
\begin{equation}
\int_{0}^{\infty }d\lambda \ \lambda \ e^{-s^2\lambda ^{2}}\left[ J_{\beta
	_{j}}^{2}(\lambda r)+J_{\beta _{j}+\epsilon _{j}}^{2}(\lambda r)\right] =
\frac{1}{2s^2 }e^{-r^{2}/(2s^2 )}\left[ I_{\beta _{j}}(r^{2}/(2s^2
))+I_{\beta _{j}+\epsilon _{j}}(r^{2}/(2s^2 ))\right] \ ,  \label{Int-reg}
\end{equation}%
with $I_{\nu }(z)$ being the modified Bessel function. Introducing a new variable
$y=r^2/(2s^2)$, and defining
\begin{equation}
\mathcal{I}(q,\alpha _{0},y)=\sum_{j=\pm1/2,\cdots}I_{\beta _{j}}(y)=\sum_{n=0}^{\infty }%
\left[ I_{q(n+\alpha _{0}+1/2)-1/2}(y)+I_{q(n-\alpha _{0}+1/2)+1/2}(y)\right]
,  \label{seriesI0}
\end{equation}%
and
\begin{equation}
\sum_{j=\pm1/2,\cdots}I_{\beta _{j}+\epsilon _{j}}(y)=\mathcal{I}(q,-\alpha _{0},y)\ ,
\label{seriesI2}
\end{equation}
the FC reads,
\begin{eqnarray}
\left\langle\bar{\Psi}\Psi\right\rangle_{s}=-\frac{qm}{(2\pi r)^{2}}\int_{0}^{\infty
}dy\ e^{-y-m^{2}r^{2}/(2y)}\mathcal{J}(q,\alpha _{0},y) \  ,  \label{FC3}
\end{eqnarray}
where we have defined the function
\begin{equation}
\mathcal{J}(q,\alpha _{0},y)=\mathcal{I}(q,\alpha _{0},y)+\mathcal{I}%
(q,-\alpha _{0},y),  \label{JCal}
\end{equation}

An integral representation for the function $\mathcal{I}(q,\alpha _{0},y)$,
suitable for the extraction of the divergent part in the FC, is derived in
\cite{mello10.2}. By using that representation, for the function (\ref{JCal})
one finds the following formula:
\begin{eqnarray}
\mathcal{J}(q,\alpha _{0},y) &=&\frac{2}{q}e^{y}+\frac{4}{\pi }%
\int_{0}^{\infty }dx\,\frac{h(q,\alpha _{0},x)\sinh x}{\cosh (2qx)-\cos
	(q\pi )}e^{-y\cosh (2x)}  \notag \\
&&+\frac{4}{q}\sum_{k=1}^{p}(-1)^{k}\cos (\pi k/q)\cos (2\pi k\alpha
_{0})e^{y\cos (2\pi k/q)}\ ,  \label{Sum01}
\end{eqnarray}%
where $p$ is an integer defined by $2p\leqslant q<2p+2$ and for $1\leqslant
q<2$ the last term on the right-hand side is absent. The function in the
integrand of (\ref{Sum01}) is given by the expression
\begin{eqnarray}
h(q,\alpha _{0},x) &=&\cos \left[ q\pi \left( 1/2+\alpha _{0}\right) \right]
\sinh \left[ \left( 1-2\alpha _{0}\right) qx\right]  \notag \\
&&+\cos \left[ q\pi \left( 1/2-\alpha _{0}\right) \right] \sinh \left[
\left( 1+2\alpha _{0}\right) qx\right] \ .  \label{g0}
\end{eqnarray}
In the above development, we have used used the notation
\begin{eqnarray}
\label{alpha}
\alpha=eA_\phi/q =-\Phi/\Phi_0= n_0+\alpha_0 \  ,
\end{eqnarray}
with $n_0$ being an integer number. So we conclude that $\mathcal{J}(q,\alpha _{0},y)$ is an even function of  $\alpha _{0}$.

We can see that the first term on the right-hand side of (\ref{Sum01})
corresponds the contribution to FC independent of $\alpha_{0}$ and $q$. In fact it
is the Minkowski spacetime contribution and in the absence of magnetic
flux. This term provides a divergent result. In order to obtain a finite and well defined FC, the standard renormalization procedure adopted in this locally flat spacetime corresponds to subtract from the complete expression the Minkowski spacetime contribution. In this way, we shall discard the exponential term
in \eqref{Sum01}. The other terms provide contributions to the FC due to the
magnetic flux and nontrivial topology of the straight cosmic string. These
terms are finite and do not require any renormalization procedure. 
Substituting (\ref{Sum01}) into (\ref{FC3}), the integrals over the variable
$y$ are evaluated with the help of formula from \cite{Grad}, and the final
result for the renormalized FC is written as:
\begin{eqnarray}
\langle \bar{\Psi}\Psi \rangle _{s}^{\mathrm{ren}} &=&-\frac{2m^{3}}{\pi ^{2}%
}\left[ \sum_{k=1}^{p}(-1)^{k}\cos (\pi k/q)\cos (2\pi k\alpha
_{0})f_{1}(2mrs_{k})\right.  \notag \\
&&+\left. \frac{q}{\pi }\int_{0}^{\infty }dx\frac{h(q,\alpha _{0},x)\sinh x}{%
	\cosh (2qx)-\cos (q\pi )}f_{1}(2mr\cosh x)\right] \ .  \label{FC4}
\end{eqnarray}%
Here we have introduced the notations
\begin{equation}
f_{\nu }(x)=K_{\nu }(x)/x^{\nu },\;s_{k}=\sin (\pi k/q),  \label{sk}
\end{equation}%
with $K_{\nu }(z)$ being the Macdonald function. For $q=1$, Eq. (\ref{FC4}) provides the FC induced by the magnetic flux in Minkowski spacetime.

\subsection{Core-induced fermionic condensate}
\label{FC_c}
As to $\left\langle\bar{\Psi}\Psi\right\rangle_{c}$, given by \eqref{condesate4},  it is finite and does not require any regularization procedure. To evaluate
this contribution we proceed as follows: in the complex $\lambda$ plane we rotate the integration
contour by the angle $\pi/2$ for $l=1$  and by the angle $-\pi/2$ for $l=2$. Moreover, it shown in \cite{mikael} that the coefficient ${\cal{V}}^{(i)}_j(\lambda, a)$ in \eqref{Z.Bessel}
satisfy the relation below:\footnote{ In fact the relation \eqref{Rel-V} is satisfied for all radial functions associated with the three configurations of magnetic field.}
\begin{equation}
\label{Rel-V}
{\cal{V}}^{(i)}_j(\pm i\lambda, a)=\pm i{\rm Im}\{{\cal{V}}^{(i)}_j( i\lambda, a)\}  \  .
\end{equation}
 In this way, by using \eqref{Rel-V} and the well known relations
involving the Hankel function of imaginary argument with the modified Bessel functions \cite{Abramo}, we can develop this calculation. The integral over the segments $(0, \ i\sqrt{m^2+k^2})$ and $(0, \ -i\sqrt{m^2+k^2})$ are canceled. In the remaining integral over the imaginary axis we introduce the modified Bessel functions. Moreover, writing imaginary integral
variable by $\lambda=\pm iz$, the core-induced fermionic condensate reads,
\begin{eqnarray}
\label{condensate5}
\left\langle\bar{\Psi}\Psi\right\rangle_{c} = \frac{q m}{2\pi^3}\int^{\infty}_{-\infty}{dk}\int^{\infty}_{\sqrt{k^2+m^2}}
dz \frac{z}{\sqrt{z^2-k^2-m^2}} \nonumber \\ \sum_{j}\left(K^{2}_{\tilde{\beta}_j}(z r)-K^{2}_{\beta_j}(z r)\right)F^{(i)}_{j}(z a) \ ,  
\end{eqnarray}
where we use the notation
\begin{eqnarray}
\label{F-ratio}
F^{(i)}_{j}(z a)= \frac{I_{\beta_j+\epsilon_j}(z a)-
\mbox{Im}[{\cal{V}}_j^{(i)}(iz/a, a)]I_{\beta_j}(z a)}{K_{\beta_j+\epsilon_j}(z a)+
\mbox{Im}[{\cal{V}}_j^{(i)}(iz/a, a)]K_{\beta_j}(z a)} \  .
\end{eqnarray}
It can be shown that switching off the magnetic field, i.e., taking $\alpha=0$, $F^{(i)}_{j}(z a)$ vanishes.
 
After a convenient coordinate transformations we rewrite \eqref{condensate5} as
\begin{eqnarray}
\label{condensate6}
\left\langle\bar{\Psi}\Psi\right\rangle_{c} = \frac{q m}{2\pi^2 r^2}\int^{\infty}_{mr}{dz}z \sum_{j}\left(K^{2}_{\tilde{\beta}_j}(z)-K^{2}_{\beta_j}(z)\right)F^{(i)}_{j}\left(z \frac{a}{r}\right) \ .
\end{eqnarray}

Before to start the explicit numerical analysis related to the core-induced 
fermionic condensate, let us now evaluate the behavior of it at large distance
from the core, i.e., let us consider the case where $mr>>ma$.  In order to develop this analysis, we rewrite the modify Bessel functions in \eqref{condensate6} in terms of their asymptotic forms.\footnote{In fact to obtain a non-vanishing result for the integrand of \eqref{condensate6}  we had to use the asymptotic expansion of the modified Bessel function until the second order.} So, we can approximate the integrand of \eqref{condensate6}  and consequently obtain an approximated expression for the FC:
\begin{equation}
\label{condensate7}
\left\langle\bar{\Psi}\Psi\right\rangle_{c} \approx \frac{q^2m}{4\pi r^2}\int^{\infty}_{mr}{dz}\frac{e^{-2x}}{x}\sum_{j}\epsilon_{j}|j+\alpha|F^{(i)}_{j}\left(z \frac{a}{r}\right) \ .
\end{equation}

Due to the exponential suppression of the integrand, the dominant contribution is given from the region near the lower limit of integration. 
Then the leading order contribution is,
\begin{equation}
\label{condensate8}
\left\langle\bar{\Psi}\Psi\right\rangle_{c} \approx \frac{q^2}{4\pi r^3}e^{-2mr}\sum_{j}\epsilon_{j}|j+\alpha|F^{(i)}_{j}\left(ma\right) \ .
\end{equation}
So, we can see that for massive fields and at large distances from the core, the 
induced fermionic condensate decays exponentially with $mr$. The correction term
in the above expression, depends on the specific model considered to
describe the magnetic field. Moreover, we also can obtain the dominant contribution for the above expression in the limit $ma<<1$. In order to do that we have to analyze, for each specific model of magnetic field, the behavior of the factor $F^{(i)}_{j}\left(m a\right)$ for small arguments. So, below we present, for the region inside the magnetic tube, the radial functions, $R_1^{(i)}(\lambda,r)$ and $R_2^{(i)}(\lambda,r)$:
\begin{enumerate}
\item For the first model, the cylindrical shell, we have:
\begin{eqnarray}
\label{model1}
R^{(1)}_1(r)=J_{\nu_j}(\lambda r) \  , \\ \nonumber
R^{(2)}_1(r)=\tilde{\epsilon}_j J_{\nu_j + \tilde{\epsilon}_j}(\lambda r) \  , 
\end{eqnarray}
where $\nu_j=q|j|-\frac{\tilde{\epsilon}_j}{2}$, with $\tilde{\epsilon}_j =1$ for $j>0$ and $\tilde{\epsilon}_j =-1$ for $j<0$.
\item For the second model, the magnetic field proportional to $1/r$, we have:
\begin{eqnarray}
\label{model2}
R^{(1)}_2(r)=\frac{M_{\kappa, \ \nu_j}(\xi r)}{\sqrt{r}} \  , \\ \nonumber
R^{(2)}_2(r)=C^{(2)}_j \frac{M_{\kappa, \ \nu_j + \tilde{\nu}_j}(\xi r)}{\sqrt{r}} \  ,
\end{eqnarray}
where
\begin{eqnarray}
\label{model2.1}
\xi=\frac{2}{a}\sqrt{q^2\alpha^2-\lambda^2 a^2}, \ \ \ \kappa=\frac{2 q^2 j\alpha}{\xi a}
\end{eqnarray}
and 
\begin{equation}
\label{model2.2}
C^{(2)}_j=\left\{\begin{array}{cc}
\frac{\lambda}{\xi}\frac{1}{(2q|j|+1)} ,  \  \ j>0  \ .
\\-\frac{\xi}{\lambda}(2q|j|+1), \ j<0 \ . \end{array} \right.
\end{equation}
\item For the third model, the homogeneous magnetic field, we have:
\begin{eqnarray}
\label{model3}
R^{(1)}_3(r)=\frac{M_{\kappa+\frac{1}{4}, \ \frac{\nu_j}{2}}(\tau r^2)}{r} \  , \\ \nonumber
R^{(2)}_3(r)=C^{(3)}_j \frac{M_{\kappa-\frac{1}{4}, \ \frac{\nu_j+\tilde{\epsilon}_j}{2}}(\tau r^2)}{r} \  , 
\end{eqnarray}
where
\begin{eqnarray}
\label{model3.1}
\tau=\frac{q\alpha}{a^2}, \ \ \ \kappa=\frac{\lambda^2}{4\tau}-\frac{qj}{2}
\end{eqnarray}
and 
\begin{eqnarray}
\label{model3.2}
C^{(3)}_j=\left\{\begin{array}{cc}
\frac{\lambda}{\sqrt{\tau}}\frac{1}{(2q|j|+1)} ,  \  \ j>0  \ .
\\-\frac{\sqrt{\tau}}{\lambda}(2q|j|+1), \ j<0 \ . \end{array} \right.
\end{eqnarray}
\\
\end{enumerate}

For the second and third models, the radial functions are given in terms of the Whittaker functions, $M_{\kappa,\nu}(z)$. 

In the limit $ma<<1$, the behavior of the fermionic condensate may be developed by the following way: first we change the summation on the angular moment $j$, in \eqref{condensate8}, by $n=j-1/2$.
In this way, we can use a new notation, $F^{(i)}_{n}
\equiv F^{(i)}_{j}$. We also change $n$ by $n-n_0$, 
being $n_0$ given in \eqref{alpha}. So from \eqref{condensate8} we can write,

\begin{equation}
\label{condensate8.1}
\left\langle\bar{\Psi}\Psi\right\rangle_{c} \approx \frac{q^2}{4\pi^2 r^3}e^{-2mr}\sum_{n=-\infty}^{+\infty}\epsilon_{n}|n+1/2+\alpha_{0}|F^{(i)}_{n-n_0}\left(ma\right) \ .
\end{equation}

By the above expression we can write $F^{(i)}_{n}\left(ma\right)$ as follow:  
\begin{eqnarray}
\label{Fratio1}
F^{(i)}_{n-n_0}(ma)= \frac{I_{{\tilde{\beta}_n}}(ma)-
\mbox{Im}[{\cal{V}}_{n-n_0}^{(i)}(im, a)]I_{\beta_n}(ma)}{K_{{\tilde{\beta}_n}}(m a)+
\mbox{Im}[{\cal{V}}_{n-n_0}^{(i)}(im, a)]K_{\beta_n}(ma)} \  .
\end{eqnarray}
In \eqref{Fratio1} the orders of Bessel functions are given by
\begin{eqnarray}
\label{betan}
\beta_n=q|n+1/2+\alpha_0|-\frac{1}{2}\frac{|n+1/2+\alpha_0|}{n+1/2+\alpha_0} \ , \nonumber\\
{\tilde{\beta}_n}=q|n+1/2+\alpha_0|+\frac{1}{2}\frac{|n+1/2+\alpha_0|}{n+1/2+\alpha_0} \ .
\end{eqnarray}

Expanding the Bessel functions in \eqref{Fratio1} in powers of $ma$, the dominant term is given by the smallest power. So we have two possibilities: for $0\leq\alpha_0<1/2$ this term
is given by $n=-1$, and for $-1/2<\alpha_0\leq 0$ this term is given for $n=0$. So, we have:
\begin{itemize}
\item For $\alpha_0>0$:
\begin{equation}
\label{a+}
F^{(i)}_{-1-n_0}\left(ma\right)\approx \frac{2}{\beta\Gamma^2(\beta)}
\frac{\beta+i{\cal{V}}_{-1-n_0}^{(i)}(im, a)
\left(\frac{ma}{2\beta}\right)}{\frac{\Gamma(1-\beta)}
{\Gamma(\beta)}-i{\cal{V}}_{-1-n_0}^{(i)}(im, a)
\left(\frac{ma}{2}\right)^{-2\beta+1}} \  .
\end{equation}
\item For $\alpha_{0}<0$, and
\begin{equation}
\label{a-}
F^{(i)}_{-n_0}\left(ma\right)\approx \frac{2}{\beta\Gamma^2(\beta)}\left(\frac{ma}{2}\right)^{2\beta}
\frac{1+i{\cal{V}}_{-n_0}^{(i)}(im, a)\left(\frac{2\beta}{ma}\right)}
{1-i\frac{\Gamma(\beta-1)}{\Gamma(\beta)}{\cal{V}}_{-n_0}^{(i)}(im, a)\left(\frac{ma}{2}\right)}  \  .
\end{equation}
\end{itemize}

The next steps are the calculations of the dominants contribution for
the coefficient that contains all the information about the core, ${\cal{V}}_{-1-n_0}^{(i)}(im, a)$ 
and ${\cal{V}}_{-n_0}^{(i)}(im, a)$, for the three models. This can be done by explicit 
substitution of the radial functions, $R^{(i)}_1(im, a)$ and $R^{(i)}_2(im, a)$,
into \eqref{Z.Bessel}.  So, for $ma<<1$, we find:
\begin{equation}
\label{condensate8.2}
\langle \bar{\Psi}\Psi \rangle_{c}\approx \frac{q^2}{2\pi^2 r^3 \Gamma^2(\beta)}e^{-2 m r} |1/2-|\alpha_0||\left(\frac{ma}{2}\right)^{2\beta}\frac{\beta-\chi^{(l)}}{\beta\chi^{(l)}} \  ,
\end{equation}
where
\begin{equation}
\label{beta-a}
\beta=q\left(\frac{1}{2}-|\alpha_{0}|\right)+\frac{1}{2}  \
\end{equation}
and $\chi^{(l)}$ is a parameter depending on the specific model adopted for the
magnetic field, given bellow by:
\begin{equation}
\label{models}
\chi^{(l)}=\left\{\begin{array}{ccc}
\nu=q|n_0-\frac12\frac{|\alpha_0|}{\alpha_0}|-\frac12\frac{|\alpha_0|}{\alpha_0}, \ \mbox{for the model ({\it{i}})}\\
q\alpha(q+1)\frac{M_{\frac{q}{2}\frac{|\alpha_0|}{\alpha_0},\nu}(2q\alpha)}
{M_{\frac{q}{2}\frac{|\alpha_0|}{\alpha_0},\nu+1}(2q\alpha)}, \ \mbox{for the model ({\it{ii}})}\\
\frac{\sqrt{q\alpha}}{2}(q+1)\frac{M_{-\frac{1}{2}\frac{|\alpha_0|}{\alpha_0}
\frac{q+1}{2},\frac{\nu}{2}}(\tau R^2)}{M_{-\frac{1}{2}\frac{|\alpha_0|}
{\alpha_0}\frac{q+1}{2},\frac{\nu}{2}}(\tau R^2)}, \ \mbox{for the model ({\it{iii}})} \ .
\end{array}\right.
\end{equation}

Our next analysis will be to investigate $\left\langle\bar{\Psi}\Psi \right\rangle_{c}$ near the core, i.e., $r\geqq a$, for the three different configurations of magnetic field. 

Now let us analyze the FC near the boundary. In general the fermionic condensate diverges near this region, so the dominant contributions in \eqref{condensate6} comes to large values of $|j|$. To find the leading term it is convenient to introduce a new variable $z=\beta_j x$, and use the uniform expansion for large order to the modified Bessel functions \cite{{Abramo}}. In order to make our work easier, we can see that \eqref{condensate6} is an even function of $\alpha$. This can be verified  by observing that changing $\alpha\to-\alpha$ and $j\to-j$, the order of Bessel functions changes as $\beta_j\to\beta_j+\epsilon_j$ and vice-verse. An analogue relation is obtained for  $\nu_j\to\nu_j+{\tilde{\epsilon}}_j$. In this way, $F^{(i)}_{j}(y)\to -F^{(i)}_{j}(y)$ and the integrand of \eqref{condensate6} is not changed. So, let us consider $\alpha>0$ in our analysis. In the summation over $j$, we can proceed as follow: for positive values of $j$, we have ${\tilde{\beta}}_j=\beta_j+1$ and for negative values of $j$, we may change $j\to -j$, so,  ${\tilde{\beta}}_j=\beta_j-1$; however, the factor $F^{(i)}_{j}$ changes its sign. As consequence we have only to consider the summation over positive values of $j$ and double the result obtained.
So, after introducing the new variable, $z=\beta_j x$, and by using the asymptotic expansion for larger order of the Macdonald function, the dominant term in the integrand of \eqref{condensate6} is:
\begin{eqnarray}
K^2_{\tilde{\beta_j}}(\beta_j x)-K^2_{{\beta_j}}(\beta_j x)\approx \frac\pi{2\beta_j}\frac1{\sqrt{1+x^2}}e^{-2\beta_j\eta}(e^{-2\eta}-1) \ ,
\end{eqnarray}
where $\eta=\sqrt{1+x^2}$. 

For the three models, it is necessary to find the leading term of $F^{(i)}_{j} (\beta_jx\frac ar)$
for large value of $j$. For the first model, this term is obtained by using expansions for the modified Bessel functions; as to the second and third models, we need the asymptotic expansion for the corresponding Whittaker function. After some long and  intermediate steps, we found that both leading terms coincide and are given bellow,\footnote{In fact the result \eqref{F_asymp} has bee derived in our previous paper \cite{mikael}.}
\begin{equation}
\label{F_asymp}
F_{j}^{(i)}\left(\beta_j x \frac{a}{r}\right)\approx \frac{1}{4\pi}\frac{e^{2 \beta_j \tilde{\eta}}}{\beta_j^2(1+e^{2\tilde{\eta}})} \ , \ i=1, \ 2, \ 3 \ ,
\end{equation}
where $\tilde{\eta}=\sqrt{1+x^2(a/r)^2}$.  

Because $j=n+1/2$, for large and positive values of $j$ we can approximate $\beta_j\approx qn$. For the three models, the core-induced fermionic condensate behaves as
\begin{equation}
\label{condensate12}
\left\langle\bar{\Psi}\Psi\right\rangle_{c} \approx \frac{m}{8 \pi^2 r^2}\sum_{n>1}{\frac{1}{n}}\int^{\infty}_{\frac{mr}{qn}}{dx}\frac{e^{-2\eta}-1}{1+e^{2\tilde{\eta}}}\frac{e^{-2 q n (\eta-{\tilde{\eta}})}}{\sqrt{1+x^2}} \ .
\end{equation}
Moreover, we also can take the approximation $\eta-\tilde{\eta}\approx x(1-a/r)$, that is valid for $x>>1$. Observing that the following ratio may be approximated by
\begin{equation}
\label{condratio}
\frac{e^{2\eta}-1}{1+e^{2\tilde{\eta}}}\approx-2 x \left(1-\frac{a}{r}\right){e^{-2\eta}} \ ,
\end{equation}
we have that the integration in \eqref{condensate12} can be approximated by
\begin{equation}
\label{condensate13}
\left\langle\bar{\Psi}\Psi\right\rangle_{c} \approx -\frac{m(1-a/r)}{4 \pi^2 r^2}\sum_{n>1}{\frac{1}{n}}\int^{\infty}_{\frac{mr}{qn}}{dx} x^2 e^{-2 q n x (1-a/r)}\ .
\end{equation}
Solving the above integration we show that
\begin{equation}
\label{condensate14}
\left\langle\bar{\Psi}\Psi\right\rangle_{c} \approx - \frac{m}{16 \pi^2 q^3 r^2}\frac{e^{-2 mr (1-a/r)}}{(1-a/r)^2} \sum_{n>1}{\frac{1}{n^4}}\ .
\end{equation}
It means that the core-induced fermionic condensate diverges near the boundary, and its divergence is proportional to
\begin{equation}
\label{condensate15}
\left\langle\bar{\Psi}\Psi\right\rangle_{c} \approx - \frac{m}{1440 q^3 r^2}\frac{\pi^2}{(1-a/r)^2} \ .
\end{equation}
Because the zero-thickness FC, $\left\langle\bar{\Psi}\Psi\right\rangle_{s}^{ren}$, is finite in this region, we conclude that the total FC, Eq. \eqref{condesate2}, is dominated by the core-induced contribution.

\begin{figure}[h]
	\centering
	\includegraphics[width=0.4\textwidth]{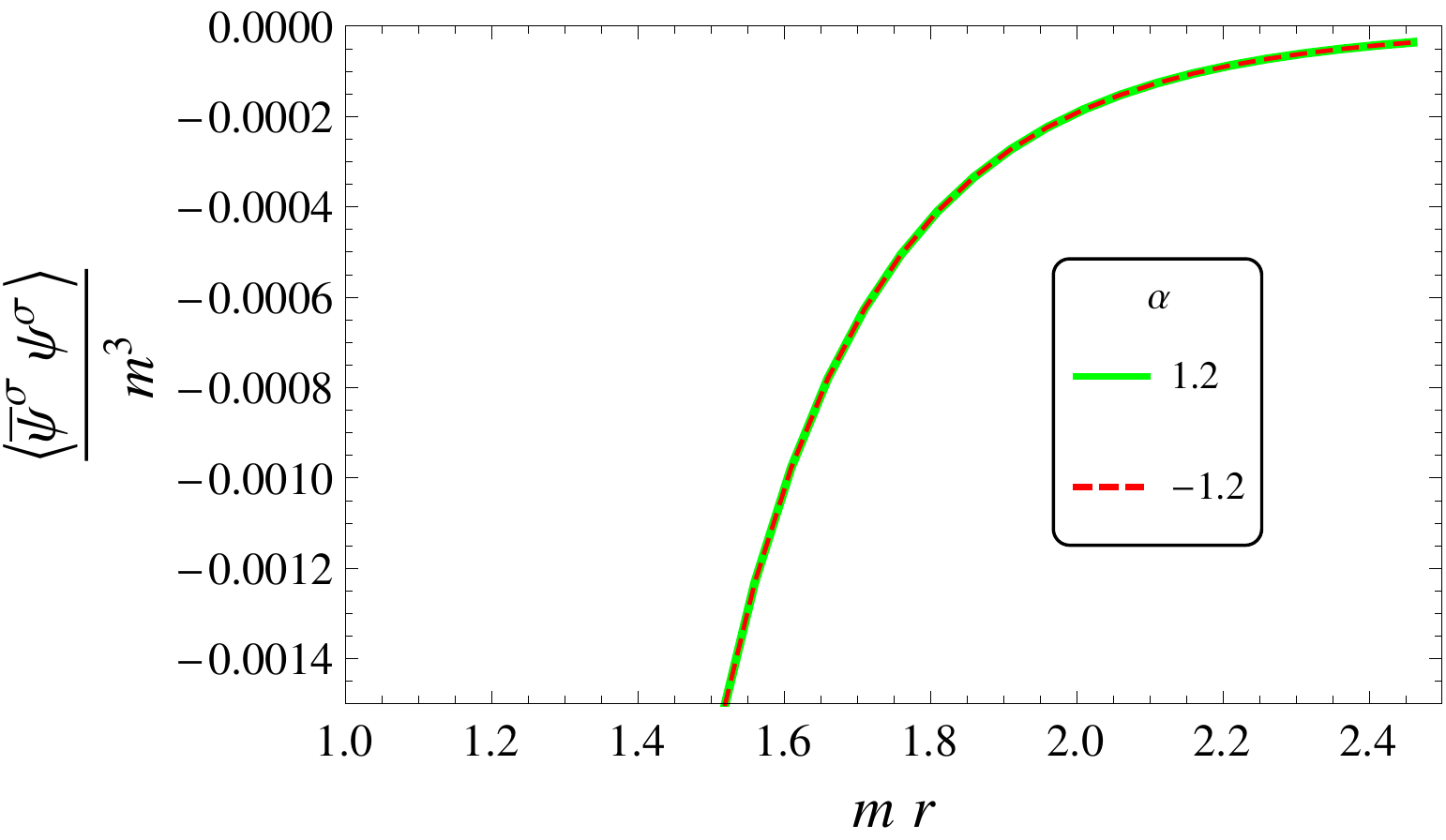}
	\includegraphics[width=0.4\textwidth]{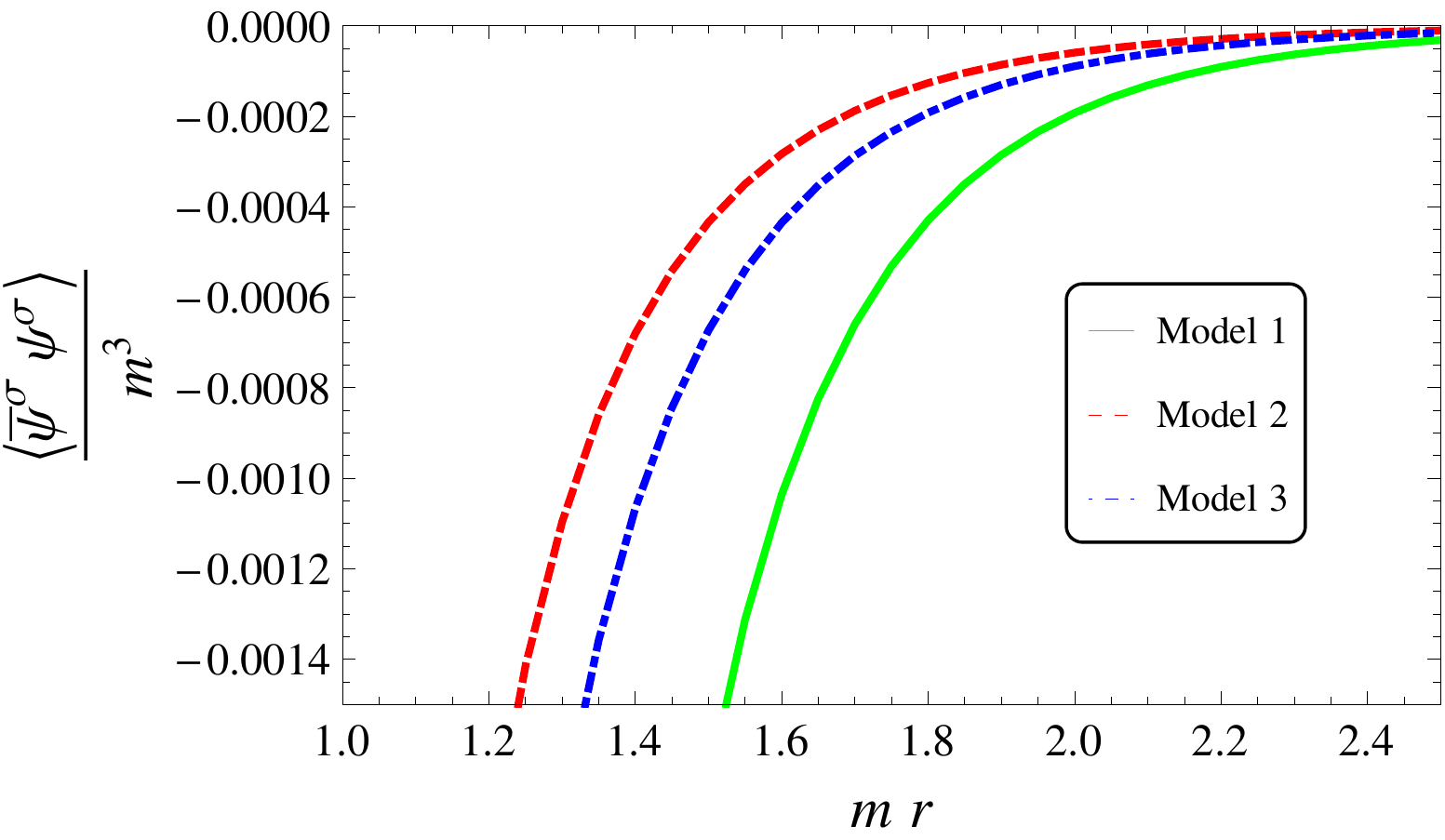}
	\caption{The core-induced fermionic condensate is plotted, in units of $``m^3"$, as a function of $mr$ for the values $q=2.0$ and $ma=1$. In the left plot, we consider the fermionic condensate induced by the magnetic field configuration of the first model, taking $\alpha=\pm1.2$. In the right plot, we compare the intensity of the core-induced fermionic condensate for the three different models of magnetic fields considering $\alpha=1.2$}
	\label{fg1}
\end{figure}

In Fig.\ref{fg1}, we display the dependence of the core-induced fermionic condensate, $\left\langle\bar{\Psi}(x)\Psi(x)\right\rangle_{c}$, as a function of $mr$ considering $q=2$ and $ma=1$. In the left plot, we present its behavior considering only the cylindrical shell of magnetic field, taking into account positive and negative values for $\alpha$. In order to provide a better understanding about this condensate, in the right plot we exhibit its behavior as function of $mr$, for the three different models of magnetic fields considering $\alpha=1.2$. By theses plots we can infer that for a given point outside the tube, the intensity of the fermionic condensate associated with the first model presents the highest value.

\begin{figure}[h]
\centering
\includegraphics[width=0.4\textwidth]{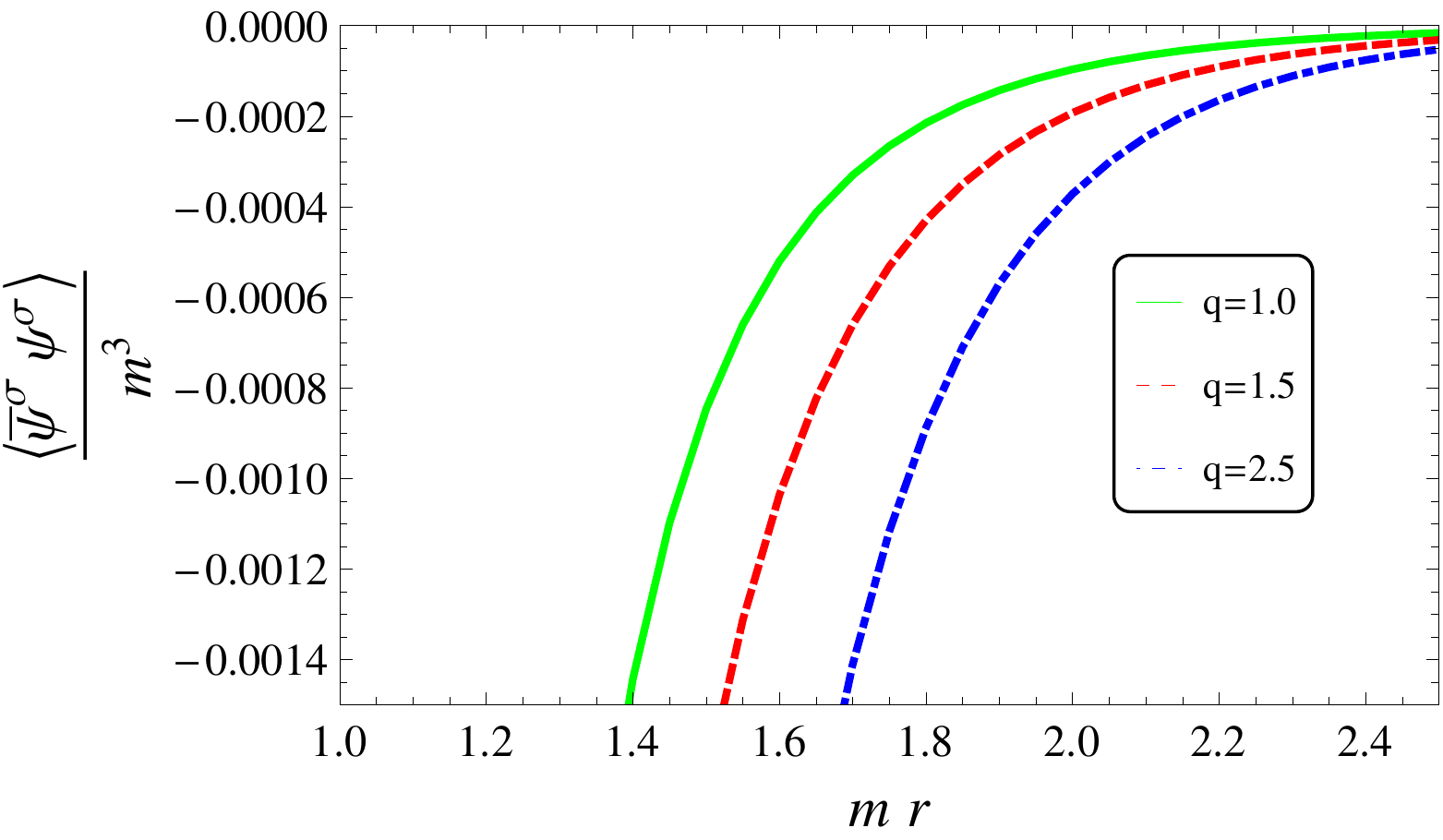}
\includegraphics[width=0.4\textwidth]{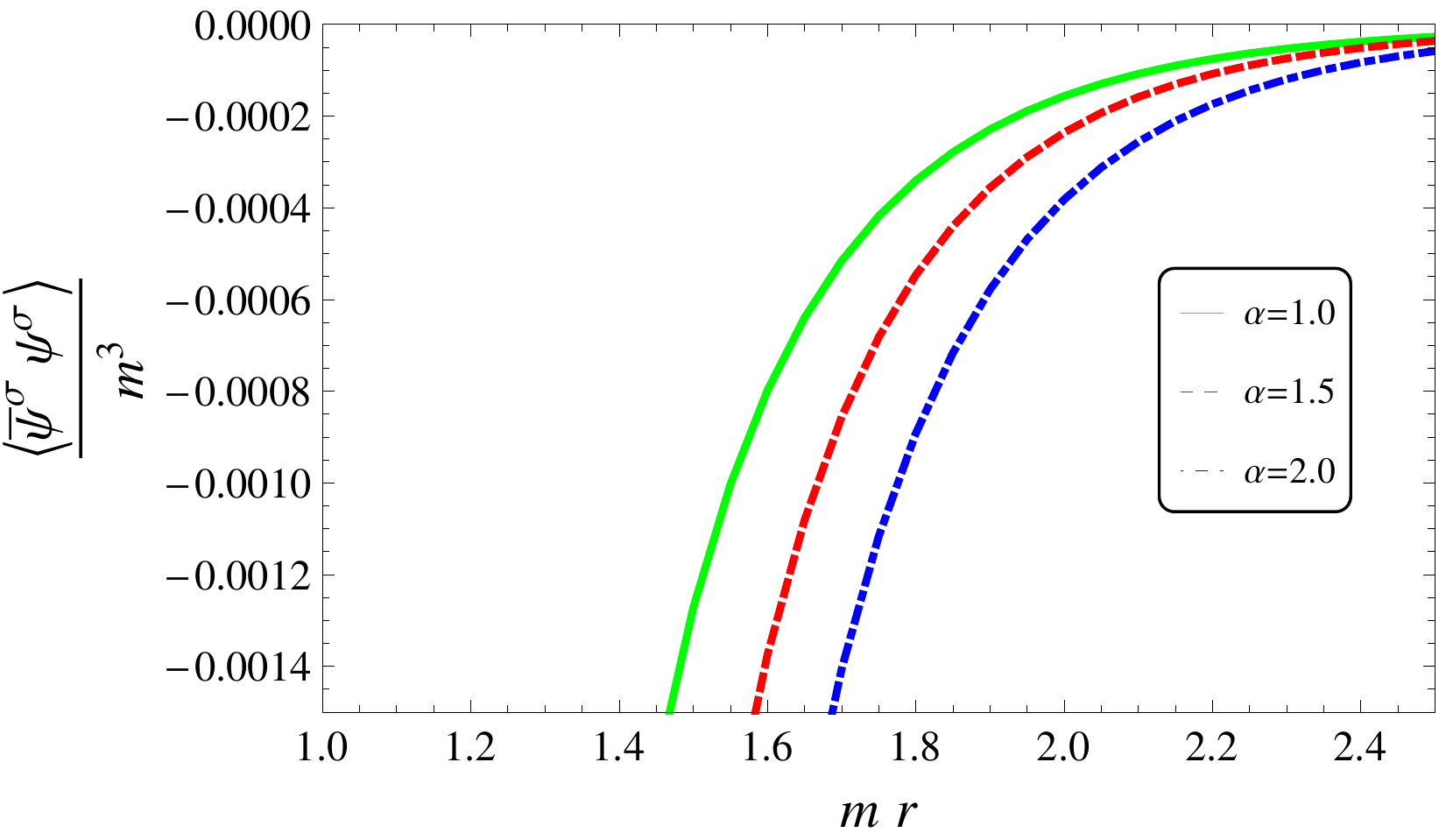}
\caption{The core-induced fermionic condensate is plotted, in units of $``m^3"$, as a function of $mr$ for the value $ma=1$. In the left plot, we consider the FC induced by the magnetic field configuration of the first model, taking $\alpha=1.2$, and $q=1, \ 1.5$ and $2.5$. In the right plot, we consider the FC induced by the magnetic field configuration of the first model, taking $q=1.5$ and $\alpha=1.0, \ 1.5$ and $2.0$. We see that for the both situations when we increases $\alpha$ or $q$ the fermionic condensate also increases.}
\label{fg2}
\end{figure}

Another two analysis that deserves to be developed are related with dependence of the core-induced fermionic condensate with the parameters which codifies the presence of the cosmic string, $q$, and the intensity of the magnetic field, $\alpha$. So. in the left plot of Fig.\ref{fg2}, we display, for the magnetic field concentrated in a cylindrical shell, the behavior of the $\left\langle\bar{\Psi}(x)\Psi(x) \right\rangle_{c}$ as a function of $mr$ for $q=1, \ 1.5$ and $2.5$ considering $\alpha=1.2$ and $ma=1$. As we can see the intensity of the FC increases with $q$. In the right plot, we display, for the same model, the behavior of the $\left\langle\bar{\Psi}(x)\Psi(x)\right\rangle_{c}$ as a function of $mr$ for $\alpha=1.0, \ 1.5$ and $2.0$ considering $q=1.5$ and $ma=1$. Also here we can infer that the FC increases with $\alpha$.

%\newpage

\section{Energy-momentum tensor}
\label{EMT}

Another important characteristic of the fermionic vacuum is the VEV of the operator energy-momentum tensor (EMT). For a charged and massive fermionic field in the presence of an electromagnetic fields, this operator is expressed as
\begin{equation}
\label{t1}
T_{\mu\nu}=\frac{i}{2}\left[\bar{\psi}\gamma_{\left(\mu\right.}\mathcal{D}_{\left.\nu\right)}\psi-(\mathcal{D}_{\left(\mu\right.}\bar{\psi})\gamma_{\left.\nu\right)}\psi\right],
\end{equation}
where $\mathcal{D}_{\mu} {\bar\psi}=\partial_{\mu}{\bar{\psi}} - ieA_{\mu}{\bar{\psi}}-{\bar{\psi}}\Gamma_{\mu}$, and the brakets in the index expression mean the symmetrization over the enclosed indexes. Similar to the case of the FC, the VEV of the energy-momentum tensor, $\left\langle 0|T_{\mu\nu}|0\right\rangle \equiv \left\langle T_{\mu\nu}\right\rangle$, can be evaluated by using the mode-sum formula.
\begin{equation}
\label{t2}
\left\langle T_{\mu\nu}\right\rangle=\frac{i}{2}\sum_{\sigma}{\left[\bar{\psi}_{\sigma}^{(-)}\gamma_{\left(\mu\right.}\mathcal{D}_{\left.\nu\right)}\psi_{\sigma}^{(-)}-(\mathcal{D}_{\left(\mu\right.}\bar{\psi}_{\sigma}^{(-)})\gamma_{\left.\nu\right)}\psi_{\sigma}^{(-)}\right]},
\end{equation}
with the notation \eqref{sum}. As in the case of the FC, we assume the presence of a cutoff without explicitly write it. We shall evaluate separately all components of the energy-momentum tensor.

In the system under consideration, the EMT can be decomposed as follow
\begin{equation}
\label{t3}
\left\langle T^{\mu}_{\nu}\right\rangle=\left\langle T^{\mu}_{\nu}\right\rangle_{s}+\left\langle T^{\mu}_{\nu}\right\rangle_{c}  \  ,
\end{equation}
where $\left\langle T^{\mu}_{\nu}\right\rangle_{s}$ is zero-thickness contribution to the VEV and $\left\langle T^{\mu}_{\nu}\right\rangle_{c}$ is the contribution induced by the non trivial structure of  the magnetic field.

In what follows we will present the expressions for all components of the EMT. Because their zero-thickness contributions have been analyzed in \cite{saharian14}, here we will concentrate only in the new contribution, the core-induced one, $\left\langle T^{\mu}_{\nu}\right\rangle_{c}$. Some details of the corresponding calculations are similar to those for the FC and the main steps only will be provided.  In \cite{sitenko12} the authors have calculated, for a quantum massive scalar field in a high dimensional cosmic string spacetime, the VEV of the energy-momentum tensor. Although the calculations have been  developed for  scalar field, the general results presented there are similar to the ones obtained  for fermionic case in \cite{saharian14}.

\subsection{Energy density}

Let us first consider the energy-density, $\left\langle T^{0}_{0}\right\rangle$. So, taking into account that $A_0$ and $\Gamma_{0}$ vanish and $\partial_t \psi^{(-)}_{\sigma}=iE\psi^{(-)}_{\sigma}$, we can write the energy density like
\begin{equation}
\label{t00}
\left\langle T^0_{0}\right\rangle=-\sum_{\sigma}{E\psi^{(-)\dagger}_{\sigma}\psi^{(-)}_{\sigma}}.
\end{equation}
By the using  (\ref{psi-out}) and \eqref{Ren1}, and taking into account that the summation over $s$, we get the following equation
\begin{equation}
\label{t00.1}
\left\langle T^0_{0}\right\rangle=-\frac{q}{4\pi^2}\sum_{j}{\int^{\infty}_{-\infty}{dk}\int^{\infty}_{0}{d\lambda\lambda}E[g^{2}_{\tilde{\beta}_{j}}(\lambda a, \lambda r)+g^{2}_{\beta_{j}}(\lambda a, \lambda r)}]  \   .
\end{equation} 

So, developing the above equation, according to \eqref{t3}, we may show that
\begin{equation}
\label{t00.s}
\left\langle T^0_{0}\right\rangle_s=-\frac{q}{4\pi^2}\sum_{j}{\int^{\infty}_{-\infty}{dk}\int^{\infty}_{0}{d\lambda\lambda}E[J^{2}_{\tilde{\beta}_{j}}(\lambda r)+J^{2}_{\beta_{j}}(\lambda r)}]
\end{equation} 
and
\begin{equation}
\label{t00.c}
\left\langle T^0_{0}\right\rangle_c=-\frac{q}{4\pi^2 r^2}\int^{\infty}_{mr}{dz z \left(\frac{z^2}{r^2}-m^2\right)\sum_{j}{[K^{2}_{\tilde{\beta}_{j}}(z)-K^{2}_{\beta_{j}}(z)}]F^{(i)}_{j}\left(z\frac{a}{r}\right)} \  ,
\end{equation}
where $F^{(i)}_{j}$ is given by \eqref{F-ratio}.

\subsection{Radial stress}

Our next step is the evaluation of the radial stress, $\left\langle T^r_{r}\right\rangle$. Taking into account that $A_r = \Gamma_r = 0$ in the general definition of the covariant derivative we have
\begin{equation}
\label{trr}
\left\langle T^r_r\right\rangle=\frac{i}{2} \sum_{\sigma}{[\bar{\psi}^{(-)}_{\sigma}\gamma^r \partial_r \psi^{(-)}_{\sigma}-(\partial_r \bar{\psi}^{(-)}_{\sigma})\gamma^r \psi^{(-)}_{\sigma}]} \  .
\end{equation}
Substituting the negative-energy fermionic mode into above expression we get
\begin{equation}
\label{trr1}
\left\langle  T^r_{r}\right\rangle=-\frac{q}{4\pi^2} \int^{\infty}_{-\infty}{dk}\int^{\infty}_{0}{d\lambda \frac{\lambda^3}{E}}\sum_{j}{\left[{g'}_{\beta_j}(\lambda a,\lambda r)g_{\tilde{\beta}_j}(\lambda a,\lambda r)-{g'}_{\tilde{\beta}_j}(\lambda a,\lambda r)g_{\beta_j}(\lambda a,\lambda r)\right]}  \   ,
\end{equation}
where the prime means the derivative with respect to the argument $\lambda r$ of the Bessel function in the definition of the $g_\beta$ functions, Eq.s \eqref{g-beta} and \eqref{g-beta1}. After some intermediate steps, we can show that $\left\langle T^r_{r}\right\rangle$ can be written as a summation over two contributions. The first one is, 
\begin{equation}
\label{trr.1}
\left\langle T^r_{r}\right\rangle_s=\frac{q}{4\pi^2} \int^{\infty}_{-\infty}{dk}\int^{\infty}_{0}{d\lambda \frac{\lambda^3}{E}}\sum_{j}{S_{j}(\lambda r)},
\end{equation}
where we have introduced the notation 
\begin{equation}
\label{sfunction}
S_{j}(z)=J^{2}_{\tilde{\beta}_j}(z)+J^{2}_{\beta_j}(z)-\frac{2\beta_j +\epsilon_j}{z}J_{\tilde{\beta}_j}(z)J_{\beta_j}(z) \  
\end{equation}
and the core-induced contribution reads,
\begin{equation}
\label{trr.c2}
\left\langle T^r_{r}\right\rangle_c=\frac{q}{2\pi^2 r^4}\int^{\infty}_{mr}{dz} z^3\sum_{j}{W_{j}(z)F^{(i)}_{j}\left(z\frac{a}{r}\right)} \  ,
\end{equation}
where we have introduced another function, $W_j (z)$, as follow
\begin{equation}
\label{wfunction}
W_{j}(z)=K^{2}_{\tilde{\beta}_j}(z)-K^{2}_{\beta_j}(z)-\frac{2\beta_j +\epsilon_j}{z}\epsilon_{j} K_{\tilde{\beta}_j}(z)K_{\beta_j}(z).
\end{equation}

\subsection{Azimuthal stress}

To evaluate the VEV of the azimuthal stress, $\langle T^{\phi}_{\phi} \rangle$, we must take into account, 
\begin{eqnarray}
	A_\phi=-\frac{q\Phi}{2\pi} \  ,
\end{eqnarray}	
being $\Phi$ the magnetic flux, and 	
\begin{eqnarray}
\label{sconection1}
\Gamma_\phi = -\frac{i}{2}(1-q)\Sigma^{(3)}, \\ \nonumber
\Sigma^{(3)}={\rm diag}(\sigma_3, \ \sigma_3) \ ,
\end{eqnarray}
with $\sigma_3$ being the Paule matrix in flat spacetime. Thus, this component reads
\begin{equation}
\label{tphi}
\left\langle T^{\phi}_{\phi}\right\rangle = \frac{i}{2}\sum_{\sigma}\left[\bar{\psi}^{(-)}_{\sigma}\gamma^\phi\mathcal{D}_{\phi}\psi^{(-)}_{\sigma}-(\mathcal{D}_{\phi}\bar{\psi}^{(-)}_{\sigma})\gamma^{\phi}\psi^{(-)}_{\sigma}\right] \ .
\end{equation}

To make our development easier we can express the derivative, $\partial_\phi$, in term of the total angular momentum operator: $\partial_\phi = i{\hat J}-i\frac{q}{2}\Sigma^{(3)}$. Then we have $\mathcal{D}_\phi = i({\hat J}+eA_\phi -\frac{1}{2}\Sigma^{(3)})$. The total angular momentum operator acting on the negative energy wave function give us the following result,  ${\hat J}\psi^{(-)}_{\sigma}(x) = qj\psi^{(-)}_{\sigma}$. Moreover, we can observe that the anticommutator, $\left\{\gamma^{\phi},\Sigma^{(3)}\right\}$, which appears in this development, vanishes. Therefore, after some steps, we get
\begin{equation}
\label{tphi1}	
\langle T^{\phi}_{\phi} \rangle =- q\sum_{\sigma}{(j+\alpha)}\bar{\psi}^{(-)}_{\sigma}\gamma^{\phi}\psi^{(-)}_{\sigma} \ .
\end{equation}	
	
Now substituting the Dirac matrix $\gamma^{\phi}$ given by \eqref{2.6} and the expressions for the negative-energy wavefunction, we obtain
\begin{equation}
\label{tphi2}
\langle  T^{\phi}_{\phi}\rangle = \frac{q^2}{2\pi^2 r}\sum_{j}{\epsilon_j (j+\alpha)}\int^{\infty}_{-\infty}{dk}\int^{\infty}_{0}{d\lambda}\frac{\lambda^2}{E} g_{\beta_j}(\lambda a, \lambda r)g_{\tilde{\beta}_j}(\lambda a, \lambda r) \ .
\end{equation}
Where the summation over $s$ provides an extra multiplicative factor $2$. So, we have for the azimuthal stress the decomposition like \eqref{t3}, with the terms given by
\begin{equation}
\label{tphi3}
\langle  T^{\phi}_{\phi}\rangle_{s}=\frac{q^2}{2\pi^2 r}\sum_{j}{\epsilon_j (j+\alpha)}\int^{\infty}_{-\infty}{dk}\int^{\infty}_{0}{d\lambda}\frac{\lambda^2}{E} J_{\beta_j}(\lambda r)J_{\tilde{\beta}_j}(\lambda r) \ ,
\end{equation}
that corresponds to the zero-thickness contribution, and
\begin{equation}
\label{tphi4}
\langle T^{\phi}_{\phi}\rangle_{c}=\frac{q^2}{\pi^2 r^4}\sum_{j}{\epsilon_j (j+\alpha)}\int^{\infty}_{mr}{dz}z^2 K_{\beta_j}(z)K_{\tilde{\beta}_j}(z) F^{(i)}_{j}\left(z\frac{a}{r}\right) \ ,
\end{equation}
that corresponds the core-induced contribution. 

\subsection{Axial stress}

In the development of the axial stress, we have to take $A_z = \Gamma_z = 0$ in the covariant derivative of the field operator, in this way we have $\mathcal{D}_z\psi^{(-)}_{\sigma} = -ik\psi^{(-)}_{\sigma}$. In addition, the matrix $\gamma^z$ coincides with the standard Dirac matrix in the flat spacetime. For this component, we get
\begin{equation}
\label{tz}
\left\langle T^z_z \right\rangle = \sum_{\sigma}{k}\bar{\psi}^{(-)}_{\sigma}\gamma^z \psi^{(-)}_{\sigma} \ .
\end{equation}

Substituting the expressions for the negative-energy wavefunction and the gamma matrix $\gamma^z$ into above expression, we obtained,
\begin{equation}
\label{tz1}
\left\langle T^{z}_{z}\right\rangle_{s} = \frac{q}{4\pi^2}\int^{\infty}_{-\infty}{dk} k^2\int^{\infty}_{0}{d\lambda}\frac{\lambda}{E}\sum_{j}{\left[J^{2}_{\beta_j}(\lambda r)+J^{2}_{\tilde{\beta}_j}(\lambda r)\right]} \ ,
\end{equation}
for the zero thickness contribution, and  
\begin{equation}
\label{tz2}
\left\langle T^{z}_{z}\right\rangle_{c} =-\frac{q}{4\pi^2 r^2}\int^{\infty}_{mr}{dz}z\left(\frac{z^2}{r^2}-m^2\right)\sum_{j}{\left[K^{2}_{\tilde{\beta}_j}(z)-K^{2}_{\beta_j}(z)\right]F^{(i)}_{j}\left(z\frac{a}{r}\right)} \ ,
\end{equation}
for core-induced one in the axial stress.

In \cite{saharian14} we have shown, by convenient transformation, that the renormalized expressions for the zero-thickness contribution for energy-density coincides with the corresponding axial stress, i.e., $\left\langle T^{0}_{0}\right\rangle_{s}^{ren}=\left\langle T^{z}_{z}\right\rangle_{s}^{ren}$.
Here we  also can observe that the core-induced contributions for the energy-density, Eq. \eqref{t00.c}, coincides with the corresponding axial stress, \eqref{tz2}, revealing also a boost invariance along the $z-$axis for this system.   
 
\section{Energy-momentum tensor and vacuum properties}

In this section we want to investigate the properties and the asymptotic behavior of the VEVs found in the previous section. We are interested to prove that the EMT  obeys the conservation condition, 
\begin{equation}
\label{covariante1}
\nabla_\mu \left\langle T^\mu_\nu\right\rangle=0 \ , 
\end{equation}
which for the system under consideration is reduced to a single differential equation,
\begin{eqnarray}
\label{cons}
\partial_r(r\left\langle T^r_r) \right\rangle= \langle T^\phi_\phi \rangle \  , 
\end{eqnarray}
and the trace relation,
\begin{equation}
\label{trace}
\left\langle T^{\mu}_{\mu}\right\rangle=m\left\langle\bar{\Psi}\Psi\right\rangle \ . 
\end{equation}

It has been proved in \cite{saharian14}, that the renormalized VEV of the energy-momentum tensor associated with the idealized case, $\left\langle T^\mu_\nu\right \rangle_s$ obeys both properties. As to the core-induced contribution, Eq. \eqref{cons} is automatically satisfied by direct substitution of \eqref{trr.c2}, \eqref{wfunction} and \eqref{tphi4}. Also the trace relation, Eq. \eqref{trace}, can be easily verified by adding all diagonal components of the core-induced components of energy-momentum tensor, \eqref{t00.c}, \eqref{trr.c2}, \eqref{tphi4} and \eqref{tz2}, and  comparing with the FC given in \eqref{condensate6}.

Now in the rest of this section, we want to investigate the behavior of the components of the core-induced VEV of the energy-momentum tensor in the region near the core,  $r\approx a$, and very from it, $r>>a$. In the latter we consider massive and massless fields. In fact we shall calculate the energy-density, that coincides with the axial stress, and the sum of the radial and azimuthal stresses. In this sense we shall use the notation $\langle  T^i_i\rangle=\langle T^r_r\rangle+\langle T^\phi_\phi\rangle$. Consequently $\langle T^{\mu}_{\mu}\rangle=2\langle T^{0}_{0}\rangle+\langle  T^i_i\rangle$.

The procedure used to develop these analysis are very similar to those used in the fermionic condensate. So, we only reproduce below the final results.

\begin{itemize}
	\item For $mr>>ma$,
\end{itemize}
\begin{equation}
\label{bigger.r1}
\left\langle T^0_0 \right\rangle_c \approx -\frac{q^2}{8 \pi r^4}e^{-2mr}\sum_{j}{\epsilon_j}|j+\alpha|F^{(i)}_{j}(ma)  \  ,
\end{equation}
\begin{equation}
\label{bigger.r2}
\left\langle T^i_i \right\rangle_c \approx \frac{q^2m}{4 \pi r^3}e^{-2mr}\sum_{j}{\epsilon_j}|j+\alpha|F^{(i)}_{j}(ma) \ ,
\end{equation}
where we can see that for larges distances the above components of the energy-momentum tensor exponential decays like in the fermionic condensate.

\begin{itemize}
	\item For $r\approx a$,
\end{itemize}
\begin{equation}
\label{r.3}
\left\langle T^0_0 \right\rangle_c \approx -\frac{3}{2880  q^3}\frac{\pi^2}{r^4\left(1-\frac{a}{r}\right)^4}-\frac{3 m}{1440 q^3}\frac{\pi^2}{r^3(1-a/r)^3}-\frac{5 m^2}{2880 q^3}\frac{\pi^2}{r^2\left(1-\frac{a}{r}\right)^2}+O\left(\frac1{1-a/r}\right)  \ ,
\end{equation}
\begin{equation}
\label{r.4}
\left\langle T^i_i \right\rangle_c \approx \frac{3}{1440 q^3}\frac{\pi^2}{r^4\left(1-\frac{a}{r}\right)^4} + \frac{3 m}{720 q^3}\frac{ \pi^2}{r^3 (1-a/r)^3} + \frac{ m^2}{360 q^3}\frac{\pi^2}{r^2\left(1-\frac{a}{r}\right)^2}+O\left(\frac1{1-a/r}\right) \ .
\end{equation}
In both expressions above we have kept all the relevant sub-leading divergent terms in order to get the correct trace relation, Eq. \eqref{trace}. 

Our next analysis is to consider the VEV of the energy-momentum tensor in the limit $r>>a$ for the massless case. Because the system presents a boost invariance  along the $z-$axis, and for massless case, we have $\langle T^i_i\rangle_c= -2{\langle T^0_0\rangle_c}$. So we need only to calculate the $\langle T^0_0 \rangle_c$ in this limit:
\begin{equation}
\label{t00+}
\langle T^0_0 \rangle_c=\frac{q}{4 \pi^2 r^4}\sum_{j}\int^{\infty}_{0}{dz}z^3\left(K^2_{\tilde{\beta}_j}(z)-K^2_{\beta_j}(z)\right)F^{(i)}_j\left(z\frac{a}{r}\right) \ .
\end{equation} 

 To explore the behavior of the energy-density in this limit we just need to take the approximated expression for the function $F^{(i)}_j\left(z\frac{a}{r}\right)$ in the limit $a/r<<1$. In fact a similar analysis has been developed in the fermionic condensate. There we have obtained the corresponding expressions given by Eq.s \eqref{a+} and \eqref{a-}. So for this case we have only to change the argument $ma$ by $za/r$. The results are given below:
\begin{itemize}
	\item For $\alpha_0 > 0$
\end{itemize}
\begin{eqnarray}
\label{F-ratio3}
F^{(i)}_{-1+n_0}(z a/r)\approx \frac{2}{\beta\Gamma^2(\beta)}
\frac{\beta+i{\cal{V}}_{-1-n_0}^{(i)}(iz, a/r)
	\left(\frac{za}{2r\beta}\right)}{\frac{\Gamma(1-\beta)}
	{\Gamma(\beta)}-i{\cal{V}}_{-1-n_0}^{(i)}(iz, a/r)
	\left(\frac{za}{2r}\right)^{-2\beta+1}} \  .
\end{eqnarray}
\begin{itemize}
	\item For $\alpha_{0}<0$, and
\end{itemize}
\begin{equation}
\label{F-ratio4}
F^{(i)}_{-n_0}\left(za/r\right)\approx \frac{2}{\beta\Gamma^2(\beta)}\left(\frac{za}{2r}\right)^{2\beta}
\frac{1+i{\cal{V}}_{-n_0}^{(i)}(iz, a/r)\left(\frac{2r\beta}{za}\right)}
{1-i\frac{\Gamma(\beta-1)}{\Gamma(\beta)}{\cal{V}}_{-n_0}^{(i)}(iz, a/r)\left(\frac{za}{2r}\right)}  \  .
\end{equation}
Substituting the above expressions into \eqref{t00+}, and taking into account, for all different models, the radial function,  $R^{(i)}_{l}(iz,a/r)$, in the limit $a/r<<1$, after some intermediate steps, we get:
\begin{eqnarray}
\label{massless1}
\left\langle {T^0}_0\right\rangle \approx -\frac{q}{2^{4\beta}\pi^2 r^4}\frac{\beta-\chi^{l}}{(\frac{r}{a})^{2\beta}\beta\chi^{l}}
\frac{\Gamma(2+\beta)}{\Gamma^2 (\beta)\Gamma(4+2\beta)}\left(2\Gamma(2+\beta)\Gamma(1+2\beta)-\Gamma^2(2+2\beta)\right) \  ,
\end{eqnarray}
Where $\beta$ and $\chi^{(l)}$ are given by Eq.s \eqref{beta-a} and \eqref{models}.

So we can conclude that in the massless case and for $r>>a$  the energy-momentum tensor decays with $r^4\left(\frac{r}{a}\right)^{2\beta}$.

We finish this discussion by saying that in all limits taken, the trace relation is preserved.

After these above general descriptions about the core-induced VEV of the energy-momentum tensor, we would like to provide some additional informations which are not easily observed by the analytical expressions. So, in order to full fill this lack, in the rest of this section we will develop some numerical evaluations.

\begin{figure}[h]
\centering
\includegraphics[width=0.4\textwidth]{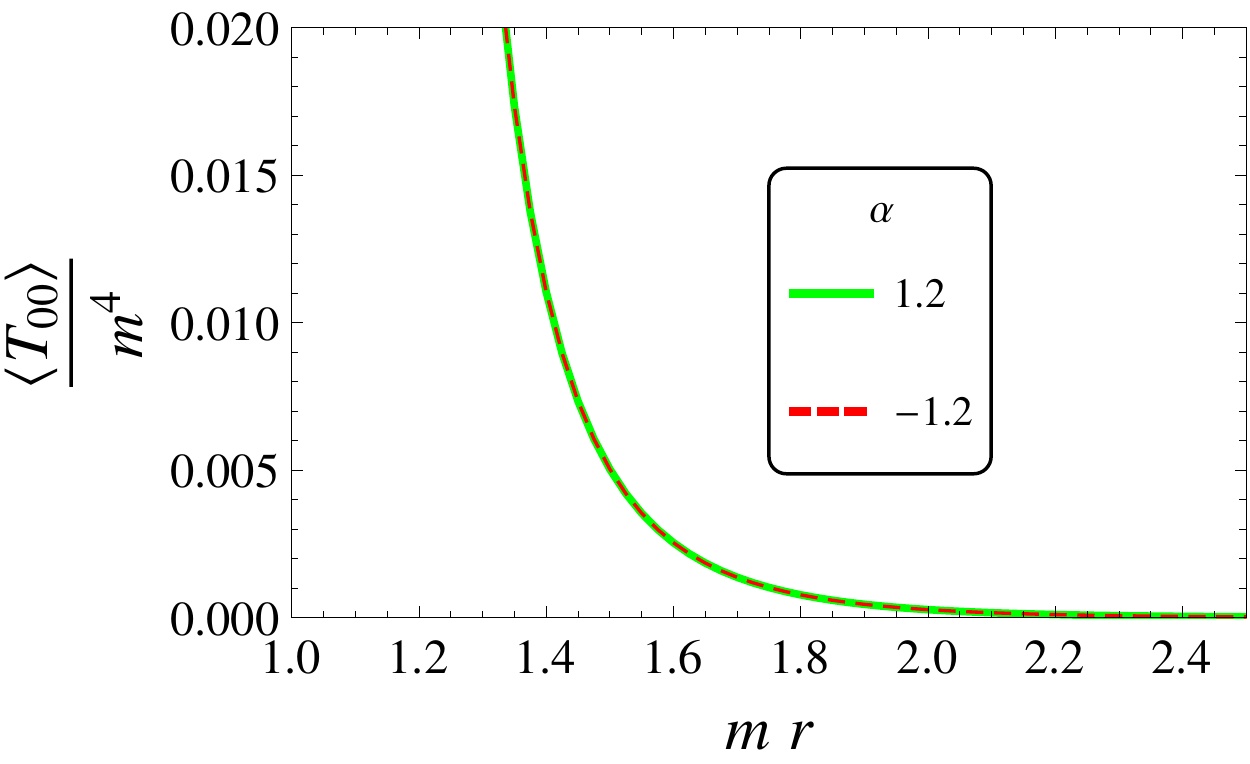}
\includegraphics[width=0.4\textwidth]{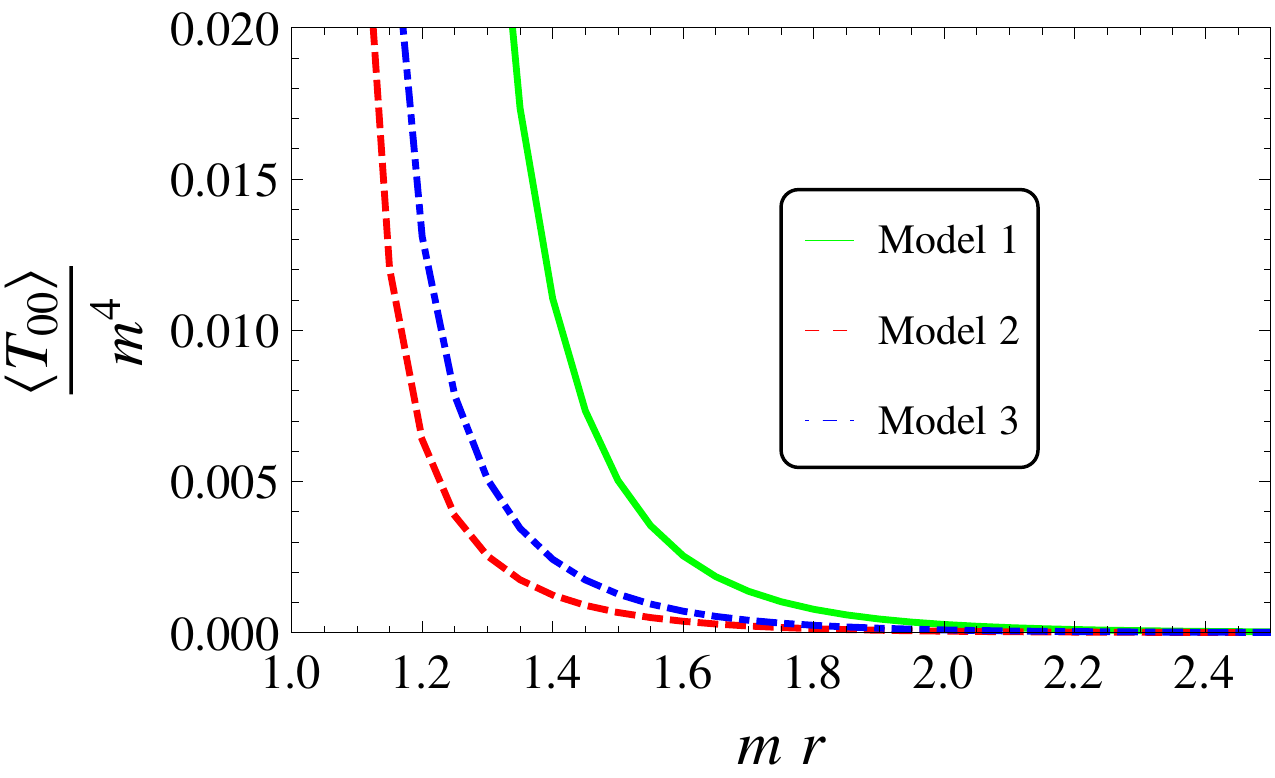}
\caption{The core-induced energy-density is plotted, in units of $``m^4"$, as a function of $mr$ for the value $ma=1$. In the left plot, we consider the magnetic field configuration of the first model, taking $\alpha=\pm 1.2$ and $q=2$. In the right plot, we consider the three different configurations of magnetic fields, taking $\alpha=1.2$ and $q=2$.}
\label{fg3}
\end{figure}

In Fig.\ref{fg3}, we display the dependence of the core-induced energy density, $\left\langle T^0_0\right\rangle_c$, as a function of $mr$ taking $q=2$ and $ma=1$. In the left plot, we present its behavior considering the cylindrical shell of magnetic field, first model, taking into account positive and negative values for $\alpha$. In the right plot, in order to provide a better understanding about the energy density, we show its behavior for the three models considering $\alpha=1.2$. So, by the right plot, we may infer that for a given point outside the tube the first model presents the higher intensity.
  
\begin{figure}[h]
\centering
\includegraphics[width=0.4\textwidth]{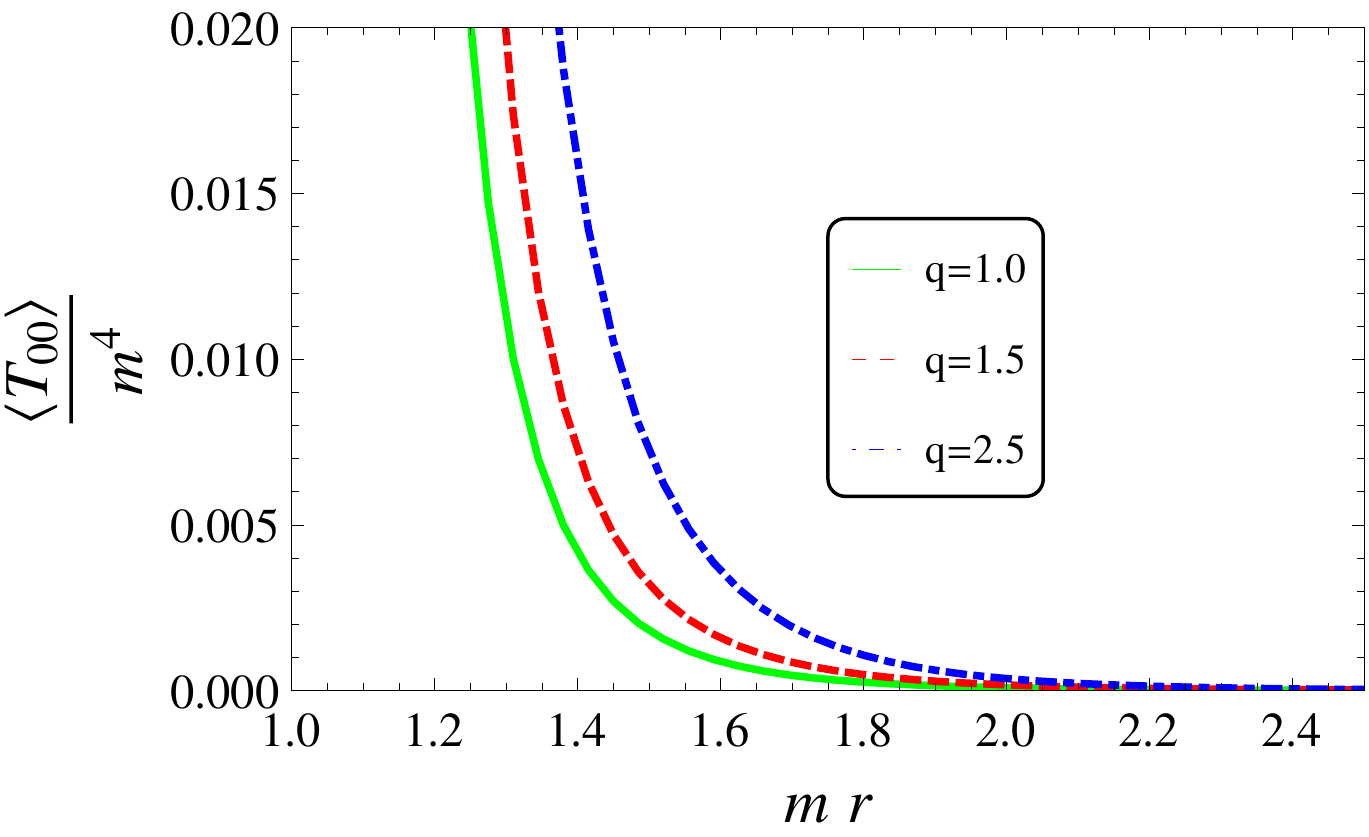}
\includegraphics[width=0.4\textwidth]{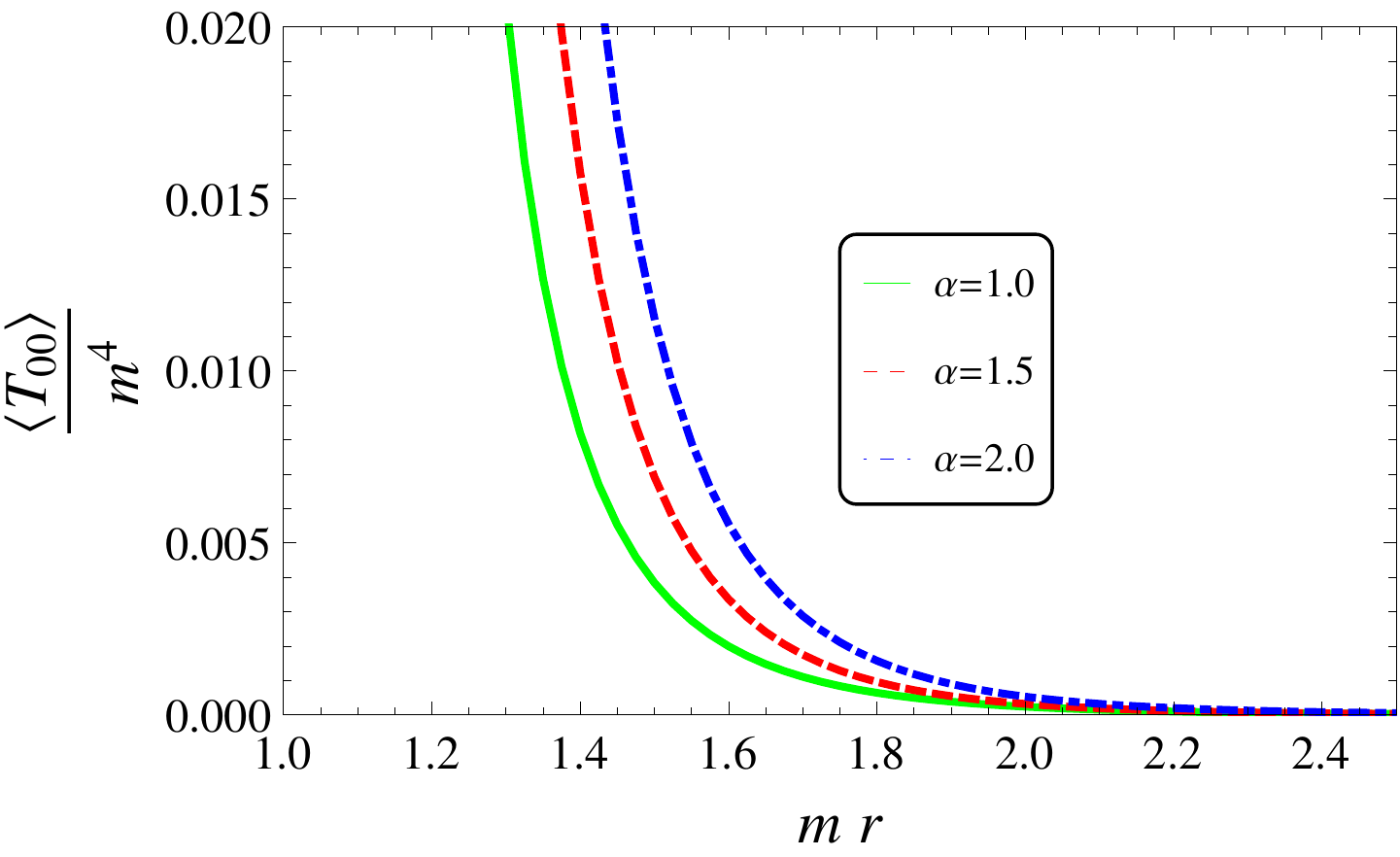}
\caption{The core-induced energy-density is plotted, in units of $``m^4"$, as a function of $mr$ for the value $ma=1$. In the left plot, we consider the core-induced energy-density taking $q=1, \ 1.5$ and $2.5$ and fixing $\alpha=1.2$. In the right plot, we consider the three different intensity to the magnetic field, taking $\alpha=1, \ 1.5$ and $2.0$ and fixing $q=2$. Both plots have been constructed by taking the magnetic configuration of the first model.}
\label{fg4}
\end{figure}

Another analysis that we develop here are related with the dependence of the core-induced energy-density with the parameters $q$, which codifies the planar angular deficit, and the parameter $\alpha$, which is related with the intensity of the magnetic field. For both analysis we concentrate in the model of the cylindrical shell only. So, in Fig.\ref{fg4} we display the behavior of the core-induced energy-density as a function of $mr$ considering the first model. In the left plot we exhibit its behavior for a fixed value of the magnetic flux, $\alpha=1.2$ and taking three different values for the angular deficit parameter, $q=1, \ 1.5$ and $2.5$; and in the right plot we exhibit the behavior of the energy-density for a fixed value of the angular deficit parameter, $q=2$, taking into account three different values of the magnetic flux, $\alpha=1, \ 1.5$ and $2$.  From both plots, we can see that by increasing the above parameters the intensity of the energy-density increases

\section{Conclusions}
\label{Concl}

In this paper we have investigated the influence of the conical topology of the spacetime, and the presence of a magnetic field of finite extension, on the vacuum polarization effects associated with massive charged fermionic field. Specifically we calculated the fermionic condensate, $\langle{\bar{\Psi}}\Psi\rangle$,  and the VEV of the energy-momentum tensor, $\langle T^\mu_\nu\rangle$, in the region outside the tube. In this analysis we adopted that the geometry of the spacetime corresponds to an idealized cosmic string everywhere, surrounded by a magnetic tube of radius $a$ for the three different models of magnetic field. The first configuration is presented by the cylindrical shell of magnetic field,  the second one is related to a field that decays with $1/r$ and the third one is presented by a homogeneous magnetic field. In order to developed these analyses, we had to construct the normalized fermionic wave-functions for the region outside the tube and calculate the fermionic condensate and the EMT by using the mode summation method, as we can see in \eqref{condesate}, \eqref{sum} and \eqref{t2}. This complete set of the fermionic wave-function was constructed by the imposition on the fermionic wave-function the continuity in the outside and inside  regions at the boundary $r=a$. After that, we got the normalized wave-function with positive- and negative-energy, given by Eq.s \eqref{psi-out}-\eqref{2.26}.

Adopting the mode sum, we have shown that the expressions found for 
the fermionic condensate and VEV of the energy-momentum tensor, are decomposed  into two distinct contributions, as we can see in \eqref{condesate2} and \eqref{t3}. The first ones, $\left\langle \bar{\Psi}\Psi\right\rangle_s$ and $\left\langle T^\mu_\nu \right\rangle_s$, depends only on the fractional part of the ratio of the total magnetic flux to the quantum one, which are consequence of  Aharonov-Bohm-like effects. The second contributions, named {\it core-induced} contributions, represented by $\left\langle \bar{\Psi}\Psi\right\rangle_c$ and $\left\langle T^\mu_\nu \right\rangle_c$, in general are not periodic functions of the magnetic flux and depends on the total magnetic flux inside the core. 

Also in this paper we have presented general analysis regarding the behavior of core-induced FC and VEV of the energy-momentum. We had observe that, considering massive fermionic fields and for large distance from the tube, $mr>>ma$, the core-induced FC and the VEV of the energy-momentum tensor exponential decays. For the FC we found $\frac{e^{-2mr}}{r^3}$ and in the same limit for the energy-density we found $\frac{e^{-2mr}}{r^4}$. We have also analyzed the core-induced FC and the VEV of the energy-momentum tensor for two other asymptotic regions of the parameters; however, in order to do that, we had to use the explicitly solutions for the radial functions inside the tube for the three different models of magnetic field. Considering points near the the magnetic tube, we have shown that the FC presents a divergent behavior as $\frac{1}{r^{2}(1-a/r)^2}$. At the same limit the spacial components of the VEV of energy-momentum tensor, $\langle T^0_0\rangle_c$ and $\langle T^i_i\rangle_c$, present a divergent behaviors given by $\frac{1}{r^{4}(1-a/r)^4}$. Because the zero-thickness contributions to the FC and VEV of the energy-momentum tensor are finite in this region, we conclude that both quantities given by, \eqref{condesate2} and \eqref{t3}, respectively, are dominated by the core-induced contributions. Moreover, we also have shown that the results found satisfy the trace relation, given by \eqref{trace}. For a massless field and at larges distances from the core, i.e., $r>>a$ we have found that the core-induced energy-density decays with $\frac{1}{r^4 (r/a)^{2\beta}}$. Because the zero-thickness contribution behaves as $\frac1{r^4}$, we conclude that in this region, the latter is the dominant contribution to the energy-density.

To finish our analysis, we have provided, by using numerical evaluation, the behavior of the core-induced FC and the VEV of the energy-momentum tensor as functions of several physical relevant quantities. In Fig.\ref{fg1} we have two plots. In the left we got an expected result: the FC is an even function of $\alpha$. In the right plot, is presented the behavior of the FC for the three models as function of $mr$. It is shown that the intensity of the FC induced by the first model presents the biggest value. In Fig.\ref{fg2}, we exhibit the behavior of the FC for the first model as a function of $mr$. In the left plot we show that for higher values of the parameter $q$, taking a fixed value of $\alpha$, the intensity of the FC increases. In the right plot we show that for higher values of the parameter $\alpha$, taking a fixed value of $q$, the intensity of the FC also increases. Following with the numerical evaluations, we can also show more two plots for the energy density. In Fig.\ref{fg3}, we have shown that, in the left plot, we got an expected result: the energy-density is an even function of $\alpha$. In the right plot we show the behavior of the energy density as function of $mr$ for the three models of magnetic fields, and we can infer that the biggest intensity corresponds to the first model. In Fig.\ref{fg4}, we present the behavior of the energy-density considering the first model only. In the the left plot we display its behavior when we increase the parameter $q$ fixing a value to $\alpha$, and in the right plot we display the behavior to energy-density when we increase the parameter $\alpha$ fixing a value to $q$. As our final remarks about the plots, we can infer that both, the FC and the energy-density, increase their intensity by increasing  the values of the parameter $q$ and $\alpha$.
	
As it was mentioned in the {\it Introduction}  there is no strong evidence of the existence of Abelian vortexes on cosmos.  However, if we are inclined to admit their existence, the interaction of these topological objects with the vacuum, would provide beautiful phenomena like induced current and energy-momentum tensor. These are the main objective of this paper. However, to obtain the complete information about these VEVs, it is necessary to have the knowledge of the inner structure of cosmic string, and unfortunately there are no closed mathematical expressions that reproduce the behavior of the magnetic field and also the core. So, the relevance of the model adopted in this paper, considering an Abelian magnetic field of finite range in a conical spacetime, is the possibility to shed some light on the analysis of induced observable by the vacuum around the cosmic string. Of course a more precise analysis requires to take into account also the interaction of the field with the string's core as it was done in \cite{Aram15}. However, this refinement can be postpone to another publication.
	
We would like to finish this paper by saying that although the analysis of a quantum field in the presence of a magnetic field confined in a cylindrical tube has been developed in the calculations of the VEVs of the energy-momentum tensor associated with the massless scalar \cite{jean03} and fermionic \cite{jean04.1} fields, this paper together with \cite{mikael} are the only ones that consider an complete analysis associated with massive fermionic field.

\textbf{Acknowledgments}: We thank Coordena\c{c}\~ao de Aperfei\c{c}oamento de Pessoal de N\'{\i}vel Superior (CAPES) and Conselho Nacional de Desenvolvimento Cient\'{\i}fico e Tecnol\'ogico (CNPq) for partial financial support.

\end{document}